\documentstyle[12pt,prd,aps,epsf,floats]{revtex}
\input psfig
\def\lsim{\mathrel{\rlap{\lower4pt\hbox{\hskip1pt$\sim$}}
    \raise1pt\hbox{$<$}}}         
\def\gsim{\mathrel{\rlap{\lower4pt\hbox{\hskip1pt$\sim$}}
    \raise1pt\hbox{$>$}}}         
\begin{document}
\begin{titlepage}

\title{SNO: Predictions for Ten  Measurable Quantities}
\author{{John N. Bahcall}\thanks{jnb@ias.edu}} 
\address{School of Natural Sciences, 
Institute for Advanced Study, Princeton, NJ 08540}
\author{Plamen I. Krastev\thanks{krastev@nucth.physics.wisc.edu}}
\address{Department of Physics, University of Wisconsin, Madison, WI 53706}
\author{Alexei Yu. Smirnov\thanks{smirnov@ictp.trieste.it}}
\address{International Center for Theoretical Physics, 34100 Trieste, Italy}
\maketitle
\vglue.4in

\begin{abstract}
We calculate the range of predicted values for $10$ quantities
that will be measured by the Sudbury Neutrino Observatory (SNO). We
use neutrino oscillation solutions (vacuum and MSW; active and sterile
neutrinos) that are globally consistent with all available neutrino
data and estimate realistic theoretical and experimental
uncertainties.  The neutral current to charged current double ratio is
predicted to be more than $9\sigma$ from the no-oscillation solution
for all of the currently favored neutrino oscillation solutions.  The
best-fit oscillation solutions predict a CC day-night rate difference
between $-0.1$\% and $+12.5$\% and a NC day-night difference $< 0.01$\%. We
present also the predicted range for the first and the second moments of the
charged current electron recoil energy spectrum, the charged current, the
neutral current, and the $\nu$-$e$ scattering rates, the seasonal dependence of the charged
current rate, and the double ratio of neutrino-electron scattering
rate to charged current rate.
\end{abstract}

\end{titlepage}


\section{Introduction}
\label{sec:introduction}

What can one learn from measurements with the Sudbury Neutrino Observatory
(SNO)~\cite{sno}? What are the most likely quantitative results for
each of the different experiments that can be carried out with SNO?
The main goal of this paper is to help answer these questions by
providing quantitative predictions for the most important diagnostic
tests of neutrino oscillations that can be performed with SNO.

SNO is not an experiment. Like LEP and Super-Kamiokande, SNO is a
series of experiments. We calculate the currently-favored range of
predictions for $10$ quantities that are affected by neutrino
oscillations and which SNO will measure.  For the impatient reader, we
list here the quantities that are sensitive to neutrino oscillations
which we investigate (definitions are given later in the text): first
and second moments of the recoil energy spectrum, the charged current
(CC), the neutral current (NC), and the neutrino-electron scattering
rates, the difference between the day and the night rates for both the
CC and the NC, the difference in the winter-summer CC rates, the
neutral current (NC) to charged current (CC) double ratio, and the
neutrino-electron scattering to CC double ratio.

The simultaneous analysis of all the SNO results, measured values and
upper limits, will be a powerful technique for constraining neutrino
oscillation parameters.  As an initial
step in this direction, we analyze the combined results for five
especially informative pairs of oscillation parameters.

\subsection{SNO reactions}
\label{subsec:snoreactions}

The SNO collaboration will study  charged current
(CC) neutrino absorption  by deuterium,

\begin{equation}
\label{eq:nuabs}
\nu_e + d \to p + p + e^-\ ,
\end{equation}
neutral current (NC) neutrino disassociation of
deuterium, 
\begin{equation}
\label{eq:nunc}
\nu_x + d \to n + p + \nu'_x  \ ,\quad (x=e,\mu,\,\tau),
\end{equation}
and neutrino-electron scattering (ES),

\begin{equation}
\label{eq:nuesc}
\nu_x + e^- \to \nu'_x + e^-\ ,\quad (x=e,\mu,\,\tau) .
\end{equation}
The energy of the recoil electrons can be measured for the CC
reaction, Eq.~(\ref{eq:nuabs}), and also for  the ES reaction,
Eq.~(\ref{eq:nuesc}).  For both these reactions, the operating energy
threshold for the recoil electrons may be of order $5$ MeV.  The
threshold for the NC reaction, Eq.~(\ref{eq:nunc}), is $2.225$ MeV.
Just as for radiochemical solar neutrino experiments, there is no
energy discrimination for the NC reaction.

The Kamiokande~\cite{kamiokande} and Super-Kamiokande
experiments~\cite{superk} have performed precision studies of solar
neutrinos using the neutrino-electron scattering reaction,
Eq.~(\ref{eq:nuesc}). SNO will be the first detector to measure
electron recoil energies as a result of neutrino-absorption,
Eq.~(\ref{eq:nuabs}). We have presented in Ref.~\cite{snoshow}
detailed predictions of what may be observed with SNO for the CC
(absorption) reaction.

If there are no neutrino oscillations, i.e., $\phi(\nu_e) = \phi({\rm
total})$, then the ratios of the event rates in the SNO detector are
calculated to be approximately in the following proportions: CC:NC:ES
= 2.05:1.00:0.19, i.e., the number of CC events is expected to
exceed the number of $\nu$-$e$ scattering events by about a factor of
$11$. Since the NC efficiency is likely to be only about a half of
either the CC or the ES efficiency~\cite{sno} and currently favored
oscillation solutions give $\phi(\nu_e) \sim \phi({\rm total})$, the
observed ratio of events in the SNO detector may actually be
reasonably close to: CC:NC:ES $\sim  2.0:0.5:0.2$.

In thinking about what SNO can do, it is useful to have in mind some
estimated event rates for a year of operation.  The Super-Kamiokande
event rate~\cite{superk} for neutrino electron scattering is $0.475$
times the event rate that is predicted by the standard solar
model~\cite{bp98}. If there are no neutrino oscillations and the total
solar neutrino flux arrives at earth in the form of $\nu_e$ with a
$^8$B neutrino flux of $0.475$ times the standard model flux, then one
expects about $4.4 \times 10^3$ CC events per year in SNO above a $5$
MeV threshold and about $1.1 \times 10^3$ NC events, while there
should only be about $415$ ES events. The above rates were calculated
for a $5$ MeV CC and ES energy threshold and for a $50$\% detection
efficiency for NC events.  For an $8$ MeV threshold, the estimated CC
rate is about $45$\% of the rate for a $5$ MeV threshold and the ES
rate is only about $28$\% of the $5$ MeV threshold rate. For the
currently favored oscillation solutions, the expected CC rates are
typically of order $80$\% of the rates cited above and the NC rates
are about a factor of two or three higher.

\subsection{What do we calculate?}
\label{subsec:whatcalculate}

In this paper, we calculate the likely range of quantities that are
measurable with SNO using a representative sample of neutrino
parameters from each of the six currently allowed $99$\% C.L.  domains
of two-flavor neutrino oscillation solutions.  In other words, we
explore what can be learned with SNO, assuming the correctness of one
of the six neutrino oscillation solutions~\cite{snoshow,bks98} that
is globally consistent at  $99$\% C.L. with all of the solar neutrino
experiments performed so far (chlorine~\cite{chlorine},
Kamiokande~\cite{kamiokande}, Super-Kamiokande~\cite{superk},
Sage~\cite{sage}, and GALLEX~\cite{gallex}).

\begin{table}[!b] 
\centering
\tightenlines
\begin{minipage}{3.5in}
\caption[]{\label{tab:bestfit} {\bf Best-fit global oscillation
parameters.} 
The differences of the squared masses are given in ${\rm eV^2}$. 
The survival probabilities that correspond to these best-fit solutions
are shown in Fig.~\ref{fig:survival} and Fig.~\ref{fig:survivecompare}. 
Results are taken from Ref.~\cite{snoshow}}
\begin{tabular}{lcc} 
Scenario&$\Delta m^2$&$\sin^2(2\theta)$\\
\noalign{\smallskip}
\hline 
\noalign{\smallskip}
 LMA&  $2.7\times10^{-5}$ &$7.9\times10^{-1}$\\
 SMA& $5.0\times10^{-6}$&$7.2\times10^{-3}$\\
 LOW& $1.0\times10^{-7}$ &$9.1\times10^{-1}$\\ 
 ${\rm VAC_S}$ & $6.5\times10^{-11}$ &$7.2\times10^{-1}$ \\
 ${\rm VAC_L}$ & $4.4\times10^{-10}$ &$9.0\times10^{-1}$ \\
 ${\rm Sterile}$ & $4.0\times10^{-6}$ &$6.6\times10^{-3}$\\
\end{tabular}
\end{minipage}
\end{table}

\begin{figure}[!ht]
\centerline{\psfig{figure=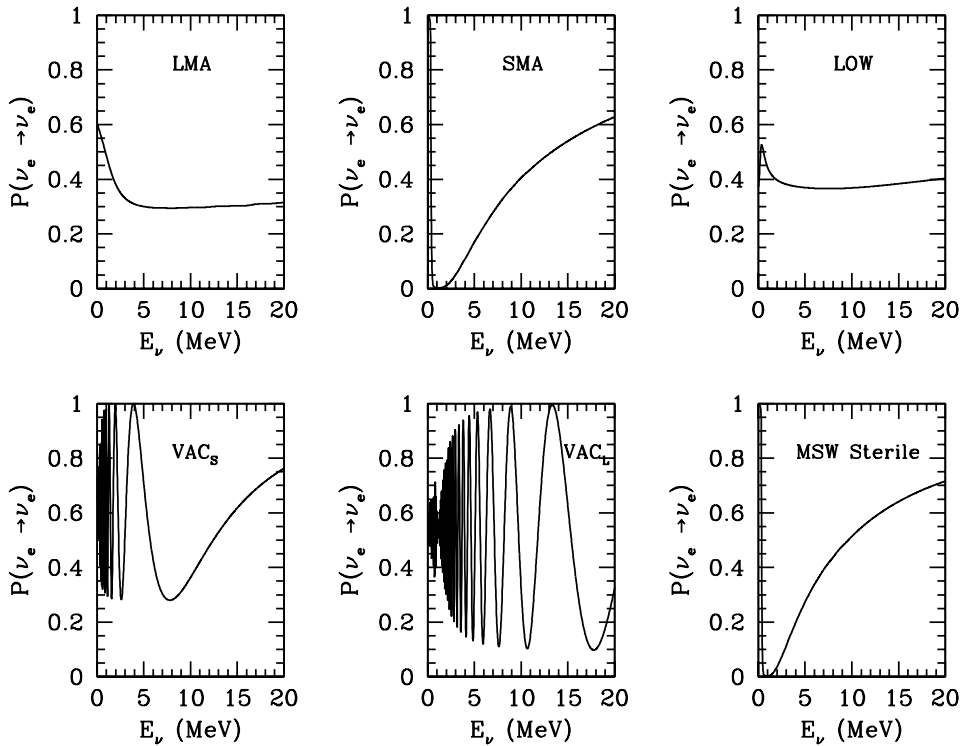,width=5in}}
\tightenlines
\caption[]{\small The survival probabilities as a function of neutrino
energy.  For each of the six best-fit globally acceptable neutrino
oscillation solutions discussed in Ref.~\cite{snoshow}, the figure
shows the survival probability for electron type neutrinos as a
function of energy. The results are averaged over one year.
\label{fig:survival}}
\end{figure}

\begin{figure}[!ht]
\centerline{\psfig{figure=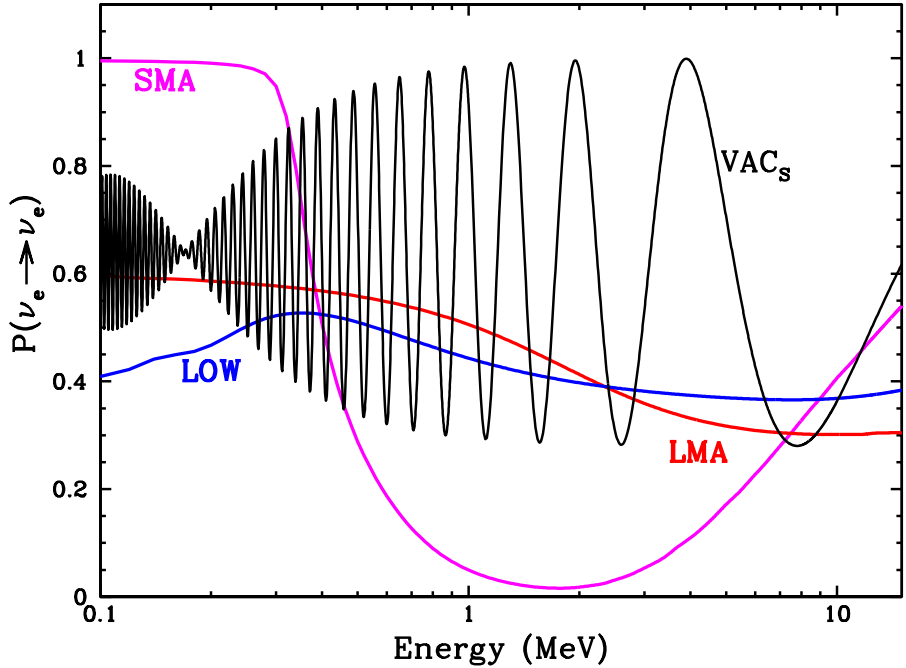,width=5in,angle=270}}
\tightenlines
\caption[]{\small Comparison of survival probabilities. For four of
the best-fit oscillation solutions, the survival probabilities are
compared using a linear energy scale. The differences in the energy
dependence of the survival probability between the high energy region,
$\sim 8$ MeV, and the low energy region, $\leq 1$ MeV, can be seen
clearly on this figure. The MSW Sterile solution has an
energy dependence similar to the SMA solution.  For clarity, we have omitted
the VAC${\rm _L}$ solution.  The parameters of the best-fit solutions
are given in Table~\ref{tab:bestfit}.
\label{fig:survivecompare}}
\end{figure}

Table~\ref{tab:bestfit} lists the mixing angles and differences of
mass squared for the six global best-fit solutions.
Figure~\ref{fig:survival} and Figure~\ref{fig:survivecompare} show the
survival probabilities of the best-fit solutions as a function of
energy{\footnote {For the MSW solutions, there are small but
perceptible differences in the computed survival probabilities which
depend upon the neutrino production probability as a function of solar
radius. The survival probabilities shown in Fig.~\ref{fig:survival}
were computed by averaging the survival probability over the $^8$B
production region in the BP98 model~\cite{bp98}. In order to portray
more accurately the behavior at low energy, the survival probabilities
for Fig.~\ref{fig:survivecompare} were computed by averaging over the
$p-p$ production region in the BP98 model.}}

For each measurable quantity $i$, we express our predictions based
upon neutrino oscillation models in terms of the value predicted by an
oscillation scenario divided by the value predicted by the combined
standard electroweak model and the standard solar model. Thus for each
measured quantity, $i$ (like CC or NC event rate), we evaluate the
expected range of the reduced quantity $[i]$

\begin{equation}
[i] ~\equiv~ \frac{{\rm (Observed~Value)_i}}{{\rm (Standard~Model~Value)_i}}.
\label{eq:defnratio}
\end{equation}
The reduced quantity $[i]$ is by construction independent of
the absolute value of the solar neutrino flux, which is used in
calculating both the numerator and denominator of
Eq.~(\ref{eq:defnratio}). 

What fluxes are used in calculating the predicted rates (e.g., for
charged current or electron-neutrino interactions) implied by
different neutrino scenarios? We determine the best-fit ratio of the
observed neutrino flux to the standard model flux by fitting to the
Super-Kamiokande rate and observed recoil electron spectrum. The
procedure is described in Sec.~\ref{sec:shapespectrum} following
Eq.~(\ref{eq:defrofe}) and Eq.~(\ref{eq:definitionsfs}) (see
especially the definition of $f(^8B)$).  We have not included explicit
uncertainties in determining $f(^8B)$ for a given pair of neutrino
variables, $\Delta m^2$ and $\sin^2 2\theta$, but we instead have
allowed $f(^8B)$ to range over the wide set of values obtained by
applying our best-fit procedure at each point in the currently-allowed
neutrino-parameter space.

As we shall see, the most
powerful diagnostics of neutrino oscillations are formed by
considering the reduced double ratio of two measurable quantities, $i$
and $j$, as follows:
\begin{equation}
{[i] \over [j]} ~\equiv~ \frac{{\rm
(Observed~Value)_i/(Standard~Model~Value)_i }}{{\rm
(Observed~Value)_j/(Standard~Model~Value)_j }} .
\label{eq:defndoubleratio}
\end{equation}
For example, the reduced double ratio of NC to CC rates is not only
independent of the absolute flux of the solar neutrinos but is also
insensitive to some experimental and
theoretical uncertainties that are important in interpreting the
separate [NC] and [CC] rates.

We describe how we evaluate the
uncertainties in Sec.~\ref{sec:uncertainties}.
All of the calculated departures from the standard model expectations
are small except for the double ratio of NC to CC,
[NC]/[CC]. Therefore, the theoretical and the experimental
uncertainties are important. 

We present in Sec.~\ref{sec:shapespectrum} the results predicted by
the six oscillation solutions for the first and second moments of the
shape of the CC recoil electron energy distribution.  We summarize in
Sec.~\ref{sec:ccrate} the principal predictions for the CC rate, in
Sec.~\ref{sec:ncrate} the predictions for the neutral current rate,
and in Sec.~\ref{sec:esc} the predictions for the neutrino-electron
scattering rate.  We then calculate the detailed predictions of the
most important double ratios, the NC to charged current ratio,
[NC]/[CC], in Sec.~\ref{sec:ncovercc} and the neutrino-electron
scattering to CC ratio, [ES]/[CC], in Sec.~\ref{sec:esctocc}.

Up to this point in the paper, i.e., through
Sec.~\ref{sec:esctocc}, we only discuss time-averaged
quantities. In Sec.~\ref{sec:daynight}, we present the predictions
for the CC of the difference between the event rate observed at night
and the event rate observed during the day. For the NC rate, there is
also a small difference predicted between the night rate and the day
rate if the MSW Sterile solution is correct. We analyze in
Sec.~\ref{sec:seasonal} the seasonal effects in the CC rate.

Section~\ref{sec:smokinggun} is a pairwise exploration of the
discriminatory power gained by analyzing simultaneously the
predictions and the observations of different smoking-gun indicators
of neutrino oscillations. We consider in this section the joint analysis
of variables like [NC]/[CC] versus the first moment of the CC energy
spectrum, the day-night difference, or the neutrino-electron
scattering rate. In Sec.~\ref{sec:discussion}, we summarize and
discuss our principal conclusions. Since we evaluate so many different
effects, we give in Sec.~\ref{sec:mostimportant} our personal list
of our top four conclusions.

\subsection{How should this paper be read?}
\label{subsec:howread}

We recommend that the reader begin by looking at the figures, which
give a feeling for the variety and the size of the various quantities
that can be measured with SNO. Then we suggest that the reader jump
directly to the end of the paper. The main results of the paper are
presented in this concluding section; the summary given in
Sec.~\ref{sec:discussion} can be used as a menu to guide the reader
to the detailed analyses that are of greatest interest to him or to
her.

This is the fifth in a series of papers that we have written on the
potential of the Sudbury Neutrino Observatory for determining the
properties of neutrino oscillations. The reader interested in details
of the analysis may wish to consult these earlier
works~\cite{snoshow,howwell,bl,bkl97}, which also provide a
historical perspective from which the robustness of the predictions
can be judged. The present paper is distinguished from its
predecessors mainly in the specificity of the predictions
(representative $99$\% C.L. predictions for each of the six currently
acceptable neutrino oscillation scenarios) and in the much larger
number of measurable quantities for which we now make predictions.

Recent review articles summarize clearly the present state of neutrino
physics~\cite{altarelireview,ramond} and
neutrino oscillation experiments and
theory~\cite{bks98,bilenkyreview,fisher,gonzalez}.  Three and four
flavor solar neutrino oscillations are discussed in
Refs.~\cite{three,four} and references cited therein.  The fundamental
papers upon which all of the subsequent solar neutrino oscillation
work is based are the initial study of vacuum oscillations by Gribov
and Pontecorvo~\cite{vac} and the initial studies of matter
oscillations (MSW) by Mikheyev, Smirnov, and Wolfenstein~\cite{msw}.
In addition to the by-now conventional scenarios of oscillations into
active neutrinos, we also consider oscillations into sterile
neutrinos~\cite{bks98,four,alexeisterile,sterilebarger,sterilevalle,sterilerabi,sterilebilenky}.

\section{Estimation of uncertainties}
\label{sec:uncertainties}

In this section, we describe how we calculate the uncertainties for
different predicted quantities.  Since the interpretation of future
experimental results depends upon the assigned 
uncertainties, we present here a full
description of how we determine the errors that we use in the
remainder of the paper.

Let $X$ represent the predicted quantity of interest, which may be,
for example, the first or second moment of the recoil energy spectrum,
the neutrino-electron scattering rate, the double ratio of neutral
current to charged current rate, the double ratio of neutrino-electron
scattering to charged current rate, or the difference between the day
rate and the night rate.  The method that we adopt is the same in all
cases. We evaluate $X$ with two different assumptions about the size
or behavior of a particular input parameter (experimental or
theoretical). The different assumptions are chosen so as to represent
a definite number of standard deviations from the expected
best-estimate.  The difference between the values of $X$ calculated
for the two assumptions determines the estimated uncertainty in $X$
due to the quantity varied.

To clarify what we are doing, we illustrate the procedure with 
specific examples. 
We begin by describing  in Sec.~\ref{subsec:theoryuncertainties} how
we calculate theoretical uncertainties and then we discuss the
detector-related uncertainties in Sec.~\ref{subsec:expuncertainties}.

\subsection{Theoretical uncertainties}
\label{subsec:theoryuncertainties}

\begin{figure}[!ht]
\centerline{\psfig{figure=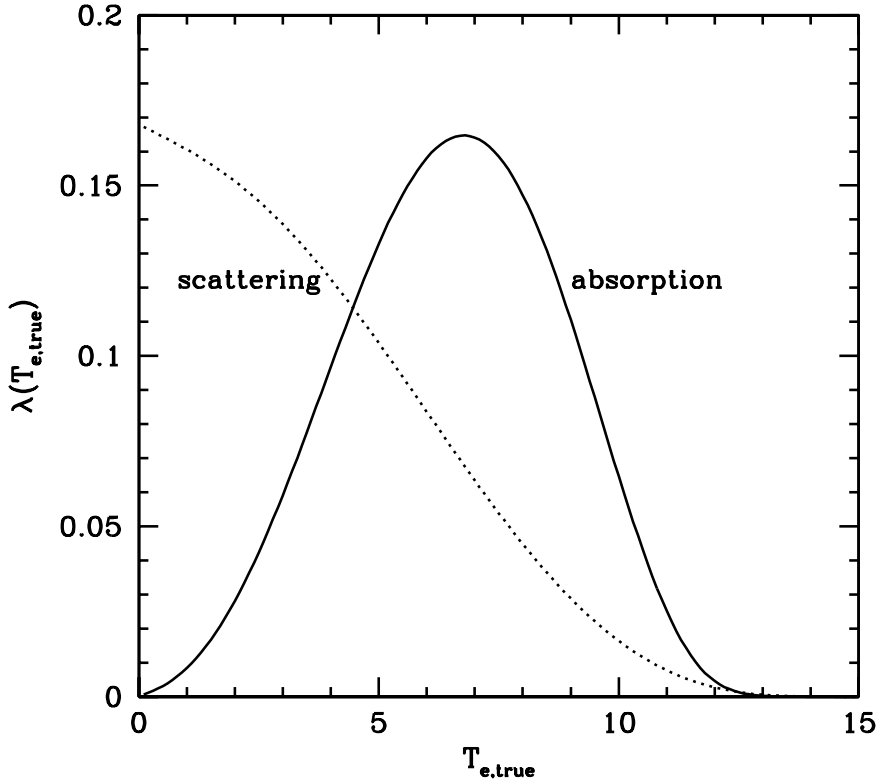,width=4.5in}}
\tightenlines
\caption[]{\small The calculated standard recoil electron energy
spectra in SNO. The figure shows the predicted shapes, $\lambda(T)$,
of the normalized recoil electron energy spectra that are obtained by
assuming that no oscillations occur and by using the standard
(undistorted) $^8$B neutrino energy spectrum. The spectra
are given as a function of the true electron kinetic
energy in MeV, $T_{\rm e,~true}$, not the apparent energy measured by
the detector. The spectra shown do not include instrumental effects
such as the finite energy resolution of the detector or uncertainties
in the absolute energy scale. The dotted curve represents the recoil
electron spectrum due to neutrino-electron scattering and the solid
curve represents the electron spectrum produced by neutrino absorption
(CC reactions) on deuterium.
\label{fig:spectra}}
\end{figure}

We discuss in this subsection the uncertainties related to the $^8$B
neutrino energy spectrum, the neutrino interaction cross sections, and
the $hep$ solar neutrino flux.  The standard shape of the $^8$B
neutrino spectrum has been determined from the best-available
experimental and theoretical information~\cite{b8spectrum}.

Figure~\ref{fig:spectra} shows the
recoil electron energy spectra calculated for neutrino-electron
scattering and for charged current (absorption) on deuterium that were
calculated using the undistorted standard $^8$B neutrino energy
spectrum.  The recoil energy spectra produced by neutrino-electron
scattering and by neutrino absorption are very different. One can
easily see from Fig.~\ref{fig:spectra} how the location of the
threshold for CC events at $5$ MeV (before the peak) or at $8$ MeV
(after the peak) could give rise to different sensitivities to
uncertainties in, e.g., the energy resolution function. This is one
of the reasons why we have calculated in the following sections
predicted values and uncertainties for two different thresholds. For
neutrino-electron  scattering, the energy distribution decreases
monotonically from low to high energies; this uniform behavior
decreases the sensitivity, relative to the absorption process, to some
uncertainties.

Two extreme deviations from the shape of the standard spectrum were
also determined using the best-available
information~\cite{b8spectrum}; these extreme shapes represent the
total effective $\pm 3\sigma$ deviations.  We calculate the quantities
$X$ that will be measured by SNO using the standard $^8$B spectrum and
the effective $3\sigma$ different spectra and determine from the
following formula the associated uncertainty due to the shape of the
$^8B$ spectrum. Thus

\begin{equation}
\sigma_X({\rm ^8B~spectrum})~=~6^{-1}
[~\vert\,X(+3\sigma ~ {\rm spectrum})
~-~X(-3\sigma ~ {\rm spectrum}) \,\vert~] .
\label{eq:specuncertain}
\end{equation}

\begin{table}[!t]
\centering
\begin{minipage}{4in}
\tightenlines
\caption[]{\label{tab:crosssections}{\bf Neutrino Interaction Cross
Sections on Deuterium.}  The table gives, in units of $10^{-42}~{\rm
cm^2}$, the neutrino charged current (CC) and neutral current (CC)
cross sections for deuterium. The cross sections calculated by
different authors (shown in column 1) have been averaged over a
standard $^8$B neutrino energy spectrum. For the CC reactions, a $5$
MeV threshold for the recoil electron energy was assumed and the
energy resolution function for SNO was approximated by
Eq.~(\ref{eq:gaussian}) and Eq.~(\ref{eq:resolution}).  The last
column gives the calculated ratios of the NC to CC ratios.} 
\begin{tabular}{lcccc} 
Authors&CC&NC&NC/CC\\
\noalign{\smallskip}
\hline
\noalign{\smallskip}
 KN\tablenote{Ref.~\cite{kn}.}&           0.979      &0.478 &0.488    \\
 YHH\tablenote{Ref.~\cite{yhh}.}&          0.923      &0.449 &0.486   \\
 EBL\tablenote{Ref.~\cite{eb} and Ref.~\cite{bl}.}& 0.889      &...      \\ 
\end{tabular}
\end{minipage}
\end{table}

Table~\ref{tab:crosssections} lists the three relatively recent
calculations for the charged current absorption cross sections on
deuterium, by Ying, Haxton, and Henley (YHH)~\cite{yhh}, by Kubodera
and Nozawa (KN)~\cite{kn}, and by Bahcall and Lisi (BL)~\cite{bl}; the
YHH and KN calculations use potential models and BL used an effective
range treatment. For the neutral current cross sections, only the YHH
and KN cross sections are available. 
If the quantity $X$ involves the
neutral current, then we define the $1\sigma$ uncertainty by
evaluating
\begin{equation}
\sigma_X({\rm NC~cross~section})~=~
\vert ~ X({\rm YHH~cross~section})
~-~X({\rm KN~cross~section}) ~\vert .
\label{eq:crossuncertain}
\end{equation}
Using the values given in Table~\ref{tab:crosssections}, we define
analogous $1\sigma$ uncertainties for quantities associated with the
CC and the double ratio, [CC]/[NC].

There is no principle of physics that enables one to set a rigorous
error estimate based upon the cross section calculations summarized in
Table~\ref{tab:crosssections}. As a practical and plausible estimate
for this paper, we have used the average of the detailed Kubodera and
Nozawa and Ying, Haxton, and Henley calculations as our best estimate
and taken the difference between these two cross sections to be an
effective $1\sigma$ uncertainty (see also the discussion by Butler and
Chen in Ref.~\cite{bc}). 
Experimental measurements with reactor anti-neutrinos are not yet
sufficiently accurate to refine and choose between different
theoretical calculations (see results in Ref.~\cite{riley}).
Had we adopted the Ellis, Bahcall, and Lisi
effective range calculation as the lower limit instead of the Ying
{\it et al.} result, we would have obtained for the CC an uncertainty
of $9.7$\% instead of $5.8$\%. Earlier, Bahcall and
Kubodera~\cite{bkubodera} estimated an effective $3\sigma$ uncertainty
of $\pm 10$\% for the neutral current cross section by calculating
cross sections with and without meson-exchange corrections, using
different sets of coupling constants, and two different nuclear
potentials.

The nuclear fusion reaction that produces $hep$ neutrinos cannot be
calculated or measured reliably~\cite{neutrinoastrophysics,bkhep}.
The shape of the electron recoil energy distribution measured by
Super-Kamiokande can be significantly influenced by the rare
high-energy $hep$
neutrinos~\cite{superk,neutrinoastrophysics,bkhep,frere}.  In this
paper, we need to evaluate the uncertainty in a variety of quantities
$X$ due to the unknown $hep$ flux.  We use the results given in the
last column of Table~3 of Ref.~\cite{snoshow}, which lists the range
of $hep$ fluxes that correspond to different oscillation solutions
that lie within the $99$\% ($\sim 2.5\sigma$) C.L. allowed range.
Given the range of listed $hep$ fluxes, we make the plausible but not
rigorous estimate that the effective $1\sigma$ uncertainty in the
$hep$ flux is currently between $0$ and $20$ times the nominal
standard estimate of $2.15 \times 10^{3}~{\rm cm^{-2} s^{-1}}$
($0.0004$ the best-estimate $^8$B flux). Therefore, we evaluate the
uncertainty due to the increase of the $hep$ flux above the nominal
standard value from the following relation
\begin{equation}
\sigma_X ~hep~{\rm flux}~=~
\vert~ X(20\times\phi (hep,{\rm~BP98}))
~-~X(0\times\phi (hep,{\rm~BP98})) ~\vert .
\label{eq:hepuncertain}
\end{equation}
The uncertainty in the $hep$ flux is asymmetric (negative fluxes are
 not physical). We calculate the lower error by replacing
 $20\times\phi (hep,{\rm~BP98})$ in Eq.~(\ref{eq:hepuncertain}) by
 $1\times\phi (hep,{\rm~BP98})$. The lower error corresponds to
 decreasing the $hep$ flux to zero. The uncertainty in the $hep$ flux
 does not dominate the error budget for any of the quantities we
 discuss. If the reader wishes to treat differently the $hep$ flux
 uncertainty, this can be done easily by using the individual
 uncertainties in Table~\ref{tab:uncertainties}.
 
For the standard solar model (SSM), the nominal ratio of the $hep$
neutrino flux to the $^8$B neutrino flux is
$4\times10^{-4}$~\cite{bp98}.  Of all the quantities we consider in
this paper, the first and second moments of the electron recoil energy
spectrum, which are discussed in Sec.~\ref{sec:shapespectrum}, are
most sensitive to the $hep$ flux. For a nominal SSM $hep$ flux, the
first moment is shifted by $3\times10^{-4}$ relative to the first
moment computed with a zero $hep$ flux. The corresponding change for
the standard deviation of the recoil energy spectrum is
$2\times10^{-3}$. Thus the $hep$ flux of the standard solar model is
of negligible importance for all of the quantities we calculate in
this paper. The $hep$ neutrino flux will have a significant effect on
the quantities computed here only if the flux exceeds the nominal
standard value by at least an order of magnitude.

Super-Kamiokande and SNO will obtain somewhat tighter constraints on
the $hep$ flux.  Measurements of the seasonal variations of the $^7$Be
flux will test vacuum neutrino scenarios that have a small $hep$ flux 
but an appreciable distortion of the Super-Kamiokande recoil energy
spectrum~\cite{berezinsky}.  Since $\sigma_X (hep{\rm~flux})$ is linearly
proportional to the allowed range of the $hep$ flux, a reduction in
the allowed range by, for example, a factor of two will reduce the
estimated value of $\sigma_X(hep{\rm~flux})$ by a factor of two.

For neutrino-electron scattering, the situation is very different. The
interaction cross sections can be calculated precisely including even
the small contributions from radiative corrections. We use in this
work the cross sections calculated in Ref.~\cite{sirlin}; the
uncertainties in these radiative corrections are negligible for our
purposes. 

In the following sections, we often quote fractional uncertainties in
percent. We define the fractional uncertainty to be the one sigma
difference divided by the average of the two values used to obtain the
error estimate. Thus the fractional uncertainty due to an increase in
the poorly known $hep$ flux is

\begin{equation}
\delta X(hep{\rm~flux} ) ~= ~100 \times { {2 \sigma_X(hep{\rm ~flux})}
\over { \vert~ X(20\times\phi(hep,{\rm~BP98}))
~+~X(0\times\phi(hep,{\rm~BP98})) ~\vert} }.
\label{eq:fractionalpercent}
\end{equation}

\subsection{Detector-related uncertainties}
\label{subsec:expuncertainties}

There are important detector-related uncertainties that can only be
determined by detailed measurements with the SNO detector and by
careful Monte Carlo simulations. 
Perhaps the most dangerous of these uncertainties are the
misidentification uncertainties, the incorrect classification of CC,
ES, and NC events.  These errors do not cancel in the double ratios
discussed later in this paper, such as [ES]/[CC] and [NC]/[CC].

SNO is a unique detector. No other detector has previously
separated the CC and the NC reactions. The only reliable way of
estimating the effects of the confusion between different neutrino
reactions is to use the full-scale Monte Carlo simulation that is
under development by the SNO collaboration.  Since the SNO
collaboration will measure the NC rate in different ways, there will
also ultimately be internal cross checks that will limit the error due
to the NC contamination of the CC and the ES rates. The ES
contamination of CC quantities like the spectrum distortion or the
day-night effect is likely to be small, since neutrino-electron
scattering is strongly peaked in the forward direction and is
estimated to be detected at only $\sim 0.1$ the CC rate. Hopefully,
misclassification errors will have only minor effects and will be well
described by the SNO Monte Carlo simulations. But, the reader should
keep in mind that the errors estimated in this paper are lower limits;
they represent errors that we can estimate quantitatively without a
large Monte Carlo simulation. We will not consider errors due to
misclassification of neutrino event in the reminder of this paper.

One can make reasonable guesses for other important experimental
uncertainties using the experience gained from previous water
Cherenkov solar neutrino experiments and preliminary Monte Carlo
studies of how the SNO detector will perform. The most important of
these quantities that need to be determined, together with their
uncertainties, are the energy resolution, the absolute energy scale,
the detector efficiencies (for energetic electrons and for neutral
current reactions), and the energy threshold for detecting CC events.
In what follows, we will adopt the preliminary characterizations for
these detector-related uncertainties used by Bahcall and
Lisi~\cite{bl}.  We now summarize briefly our specific assumptions for
these uncertainties.

Let $T_e^\prime$ be the true electron recoil kinetic energy and $T_e$ be the
kinetic energy measured by SNO. 
We adopt the resolution function $R(T_e^\prime,\,T_e)$,
\begin{equation}
R(T_e^\prime,\,T_e) = \frac{1}{\sigma(T_e^\prime)\sqrt{2\pi}}
\exp{\left[-\frac{(T_e^\prime-T_e)^2}{2\sigma(T_e^\prime)^2}\right]} ,
\label{eq:gaussian}
\end{equation}
with an energy-dependent one-sigma width $\sigma(T^\prime_e)$  given
approximately by 
\begin{equation}
\sigma(T^\prime_e) =
(1.1 \pm 0.11)\sqrt{\frac{T^\prime_e}{10 {\rm\ MeV}}}\, {\em MeV} .
\label{eq:resolution}
\end{equation}

We adopt a conservative estimate for the $1\sigma$ absolute energy
error of $\pm 100$ keV.   We will assume, for illustrative purposes,
that the threshold for detecting recoil electrons is a total energy of $5$
MeV or $8$ MeV. 

For specificity, we assume~\cite{bl} that the neutral current detection
efficiency is $0.50 \pm 0.01$ and that the detection
efficiency for recoil electrons above threshold 
is approximately $100$\%.

\subsection{Summary of uncertainties}
\label{subsec:summaryuncertainties}

In this subsection, we present a convenient table that summarizes the
estimated uncertainties for the different physical quantities that are
discussed in detail in the following sections of the paper. It may be
useful to the reader to refer back to this summary table from time-to-time
while considering the detailed presentations.

\begin{table}
\tightenlines
\caption{{\bf Fractional uncertainties in percent for some quantities
that are measurable with the SNO detector.}  Here $\delta X = 100
\sigma_X/X$.  The different quantities are defined in the following
sections: $T$ and $\sigma$ are the first and second moments,
respectively; [CC], [NC], and [ES] are the reduced charged current,
neutral current, and neutrino-electron scattering rates, respectively;
and [NC]/[CC] and [ES]/[CC] are the neutral current to charged current
and neutrino-electron scattering to charged current double ratios,
respectively.  For CC reactions, a $5$ MeV threshold was assumed for
the energy of the recoil electrons. The statistical uncertainties are
computed assuming $5000$ CC events, $1219$ NC events, and $458$ ES
events. We assumed a NC detection uncertainty of $1\sigma = 2$\%.  We
do not include uncertainties due to misclassification of neutrino
events.
\label{tab:uncertainties}}
\begin{tabular}{dddddddd}
Source& $\delta T$&$\delta \sigma$&$\delta$ [CC]&$\delta$ [NC]&$\delta$ [ES]&
$\delta$ [NC]/[CC]&$\delta$ [ES]/[CC]\\
\noalign{\smallskip}
\hline
\noalign{\smallskip}
Energy resolution&0.3&1.4&0.4&$\sim$ 0&0.1&0.4&0.3\\
Energy scale&0.8&1.1&1.5&$\sim$ 0&0.5&1.5&1.0\\
B8 spectrum&0.4&0.8&1.9&1.6&1.4&0.3&0.6\\
Cross-section&0.03&0.15&5.8&6.4&$\sim$ 0&0.4&5.8\\
Statistics\tablenotemark[1]&0.35&1.1&1.4&2.9&4.7&3.2&4.9\\
$hep$\tablenotemark[2]&0.8&0.8&2.3&2.2&1.6&0.1&0.7\\
\noalign{\medskip}
Total&1.3&2.4&6.7&7.4&5.2&3.6&7.6\\
\end{tabular}
\tablenotetext[1]{Not including background from other sources.}
\tablenotetext[2]{One $\sigma$ $hep$ upper error.}
\end{table}

Table~\ref{tab:uncertainties} shows the fractional uncertainties in
percent that we have estimated for different measurable quantities.
The quantities in the Table are defined in the following sections.
The counting uncertainties are determined assuming that a total of
$5000$ events are measured in the CC mode; the number of NC and
neutrino-electron scattering events are then about $1219$ and $458$,
respectively.  For an $8$ MeV electron energy threshold rather than
the $5$ MeV threshold used in computing Table~\ref{tab:uncertainties},
the statistical uncertainties would be increased by about $50$\% for
the purely CC quantities like $\delta T$, $\delta \sigma$, and
[CC]. For quantities related to the ES rate, the statistical
uncertainties would be increased by a factor of about $1.9$ by raising
the electron recoil energy threshold to $8$ MeV. 
The statistical error is expected, for an $8$ MeV threshold, to dominate the
uncertainty in the ES rate.

There will be additional contributions to the statistical errors from
background sources; these uncertainties can only be determined in the
future from the detailed operational characteristics of the SNO
detector. For example, the background from the CC events will increase
the estimated statistical error for the neutrino-electron scattering
events; the amount of the increase will depend upon the angular width
of the peak in the $\nu$-$e$ scattering function.  We have not
estimated uncertainties for the day-night asymmetry, $A$, defined by
Eq.~(\ref{eq:daynightdefn}) since a detailed knowledge of the detector
is required to estimate the small uncertainties in $A$.

The errors due to the uncertainties in the $hep$ flux are
asymmetric. We show in Table~\ref{tab:uncertainties} only the upper
limit uncertainties for $hep$. The lower limit uncertainties are
negligibly small for $hep$,  since the standard model flux ratio for
$hep$ to $^8$B is $0.0004$.

The actual background rates in the SNO detector are not yet known and
may differ considerably from the rates that were estimated prior to
the building of the observatory. We have therefore not attempted to
include background uncertainties, although these may well be important
for some of the quantities we calculate.

For both the CC ratio of measured to standard model rate, [CC], and
the similarly defined neutral current ratio, [NC],
Table~\ref{tab:uncertainties} shows that the absolute value of the
neutrino cross section is the dominant source of uncertainty. This
uncertainty almost entirely cancels out in the double ratio of ratios,
[NC]/[CC].  The absolute energy scale and the value of the $hep$
neutrino energy flux are the largest estimated uncertainties for the
first moment of the CC recoil energy spectrum, $\langle T\rangle$.  Counting
statistics, assuming a total of $5000$ CC events, is estimated to be
the most important uncertainty for the neutrino-electron scattering
ratio, [ES], and the neutral current to charged current double ratio
[NC]/[CC].

In the subsequent discussion, we follow the frequently adopted
practice of combining quadratically the estimated $\sigma$'s from
different sources, including theoretical errors on cross sections and
on the $hep$ flux.  If the reader prefers to estimate the total
uncertainty using a different prescription, this can easily be done
using the individual uncertainties we present.

\section{The shape of the CC electron recoil energy spectrum}
\label{sec:shapespectrum}

In this section, we make use of the fact that solar influences on the
shape of the $^8$B neutrino energy spectrum are only of order $1$ part
in $10^5$~\cite{shapeindependence}, i.e., are completely
negligible. Therefore, we compare all of the neutrino oscillation
predictions to the calculated results obtained using an undistorted
neutrino spectrum inferred from laboratory data~\cite{b8spectrum}.

Figure~\ref{fig:spectra} shows as a solid line the calculated CC
electron recoil energy spectrum that would be produced by an
undistorted $^8$B neutrino energy spectrum.  The result shown in
Fig.~\ref{fig:spectra} does not include instrumental effects such as
the energy response of the detector, but best-estimates of the
instrumental effects (see discussion in
Sec.~\ref{subsec:expuncertainties}) are included in the results
given here and in the following sections.

It is useful in thinking about the shapes of the different electron
recoil energy spectra to consider the ratio, $R(E_e)$, of the electron
energy spectrum produced by a distorted neutrino spectrum to the
spectrum that is calculated assuming a standard model neutrino energy
spectrum~\cite{spectrumratio}. We define

\begin{equation}
R(E_e) = \frac{f{\rm (^8B)} N_B(E_e) + f (hep) N_{hep}(E_e)}{N^{\rm
SSM}_B (E_e) + N^{\rm SSM}_{hep}(E_e)}
\label{eq:defrofe}
\end{equation}
where $f(^8B)$ is the ratio of the true $^8$B neutrino flux that is
created in the sun to the standard solar model $^8$B neutrino flux,
i.e., $f(^8B) = \phi(^8B)_{\rm true}/\phi(^8B)_{\rm SSM}$. The quantity
$f(hep)$ is similarly defined as the ratio of true to standard solar
model $hep$ flux. $N_B^{\rm SSM}(E_e)$ is the number of events in a 0.5~MeV energy
bin centered at $E_e$ and calculated for the SSM ${\rm ^8B}$ neutrino
flux without oscillations.  $N_B (E_e)$ is the same quantity with 
oscillations taken into account.  $N^{\rm SSM}_{hep} (E_e)$ and
$N_{hep} (E_e)$ are the corresponding numbers for the $hep$ flux.  We
have included the instrumental effects as described 
in Sec.~\ref{subsec:expuncertainties}.

We determine the best-fit value of $f({\rm ^8B})$ and $f(hep)$
for each pair of values of the oscillation
parameters, $\Delta m^2$ and $\sin^2 2\theta$, by comparing the
theoretical predictions with  the total
rate and the recoil electron energy spectrum of
Super-Kamiokande~\cite{superk,snoshow}:
\begin{equation}
f_B = f_B (\Delta m^2, \sin^2 2\theta),~~~~ 
f_{hep}  = f_{hep}(\Delta m^2, \sin^2 2\theta). 
\label{eq:definitionsfs}
\end{equation}   
The uncertainties in the values of $f({\rm ^8B})$ and $f(hep)$ are
reflected in the allowed range of  $\Delta m^2$ and $\sin^2
2\theta$, but are not included explicitly in
Table~\ref{tab:uncertainties}.

\begin{figure}[!t]
\centerline{\psfig{figure=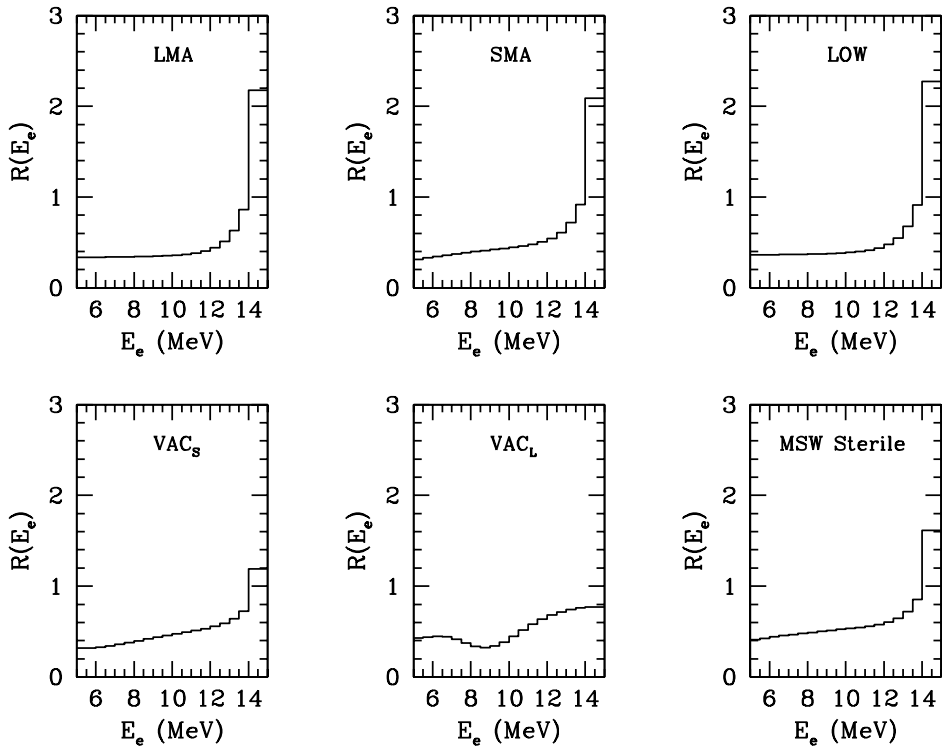,width=5in,angle=0}}
\tightenlines
\caption[]{\small The relative CC recoil energy spectra for six
oscillation solutions.  The figure shows the rates, $R(E_e)$,
predicted by different best-fit oscillation solutions as a function of
the electron recoil energy, $E_e$, divided by the rate predicted by
the standard solar model and no neutrino oscillations.  The
oscillation solutions are described in Table~\ref{tab:bestfit}. The
highest energy bin represents the average value of $R(E_e)$ for
electron energies between $14$ MeV and $20$ MeV. We use the best-fit
$hep$ flux found for each neutrino oscillation scenario, which affects
the result in the highest energy bin. The range of $hep$ fluxes is
given in Table 3 of Ref.~\cite{snoshow}.
\label{fig:sixspectra}}
\end{figure}

Figure~\ref{fig:sixspectra} shows the ratio $R(E_e)$ calculated for
the six best-fit oscillation solutions.  The values of $R(E_e)$ are
given in $0.5$ MeV bins except for the last energy bin, where we
include all CC events that produce recoil electrons with observed
energies above $14$ MeV. Only a few events (less than $1$\% of the
total number of CC events) are predicted~\cite{snoshow} to lie above
$14$ MeV since the $^8$B neutrino energy spectrum barely extends
beyond $14$ MeV and the $hep$ neutrinos, which extend up to $18.8$
MeV, are expected to be very rare.

Ultimately, SNO will measure the detailed shape of the CC recoil energy
spectrum and compare the measurements with the full predictions of
different oscillation scenarios, as illustrated in
Fig.~\ref{fig:sixspectra}. Since the neutrino oscillation parameters
are continuous variables, there are in principle an infinite number of
possible shapes to consider. However, much or most of the quantitative
information can be summarized conveniently in the first and second
moments of the recoil energy spectrum~\cite{bkl97} and we therefore
concentrate here on the lowest order moments.

Throughout this section, we use the notation of Bahcall, Krastev, and
Lisi~\cite{bkl97} (hereafter BKL97), who have defined the first and
second moments (average and variance) of the electron recoil energy
spectrum from CC interactions in SNO. The explicit expressions are
given in Eqs.~(11)--(17) of BKL97; they include the energy resolution
function of the detector [see Eq.~(\ref{eq:gaussian}) of this
paper]. Unlike BKL97, we use as our default recoil energy spectrum $5$
MeV total electron energy, rather than $5$ MeV electron kinetic
energy.  (We also calculate the moments for an $8$ MeV total electron
recoil energy.)  When we calculate for the same threshold as BKL97,
our results for the no-oscillation solution agree to about $1$ part in
$10^4$.  We use a threshold specified in terms of total electron
energy because this variable has become the standard for
experimentalists to specify their energy threshold.

We denote by a subscript of ``0'' the standard value of quantities
computed assuming no oscillations occur.  In order to compare with the
theoretical moments given here, the observed moments should be
corrected for any dependence of the detection efficiency upon energy
that is determined experimentally.

If there are no oscillations, the first moment of the CC electron
recoil kinetic energy spectrum is, for a $5$ MeV total electron energy
threshold: 

\begin{equation}
\langle T\rangle_0 = 7.422\times (1 \pm 0.013) ~{\rm MeV},
\label{eq:t0}
\end{equation}
where the estimated uncertainties ($\pm 96$ keV) have been taken from
Table~\ref{tab:uncertainties}.  The result given in Eq.~(\ref{eq:t0})
applies for a pure $^8$B neutrino spectrum. If one includes a $hep$
neutrino flux equal to the nominal standard solar model
value~\cite{bp98}, then the first moment is increased by $2$ keV to
$7.424$ MeV.  For an $8$ MeV energy threshold, $\langle T\rangle_0 =
9.117 $ for a pure $^8$B neutrino energy spectrum and is increased by
$3$ keV by adding a nominal $hep$ flux.
 
The largest estimated contributions to the quoted
error in Eq.~(\ref{eq:t0}) arise from uncertainties in the energy
scale and from the $hep$ reaction, with smaller contributions from the
width of the energy resolution function and the shape of the $^8$B
neutrino energy spectrum. The total error of the measured value is the
same, within practical accuracy, whether or not one includes the
statistical uncertainty for $5000$ events.

The first moment has the smallest estimated total error of all the
quantities tabulated in Table~\ref{tab:uncertainties}.

\begin{table} 
\centering
\tightenlines
\caption[]{\label{tab:firstmoment} {\bf The first moment, \boldmath$\langle T\rangle$, of
the electron recoil kinetic energy spectrum from CC interactions.}  If
there are no oscillations, the expected value of the first moment is
$\langle T_0\rangle = 7.422$ MeV for a 5 MeV total electron energy threshold and
$\langle T_0\rangle = 9.117$ MeV for an 8 MeV energy threshold.  The table shows
for different neutrino oscillation scenarios the absolute shift,
$\Delta T = \langle T\rangle - \langle T\rangle_0$, in keV of the first moment of the electron
recoil energy spectrum.  Results are given for both a 5 MeV and an 8
MeV threshold energy. The different columns give the best-fit shift as
well as the minimum and maximum shifts at 99\% C.L.}
\begin{tabular}{rrrrrrr} 
Scenario&${(\Delta T)}_{\rm b.f.}$&${(\Delta T)}_{\rm min}$&${(\Delta
T)}_{\rm max}$&${(\Delta T)}_{\rm b.f.}$&${(\Delta T)}_{\rm
min}$&${(\Delta T)}_{\rm max}$\\
&keV&keV&keV&keV&keV&keV\\
&5 MeV&5 MeV&5 MeV&8 MeV&8 MeV&8 MeV\\
\noalign{\smallskip}
\hline
\noalign{\smallskip}
 LMA            &  8   &$-115$  & 34 &4      &$-35$  &15 \\
 SMA            &218   &50    &341 &66      &15   &105 \\
 LOW            & 12   &$-17$  & 63 &7       &$-5$  &25 \\
  ${\rm VAC_S}$ &283   &$-80$  &576 &122       &40  &227 \\
${\rm VAC_L}$   & 21   &$-152$  &214 &236       &$-54$  &358 \\
${\rm Sterile}$ &164   &41   &265 &51       &13   &83 \\
\end{tabular}
\end{table}

Table~\ref{tab:firstmoment} presents the best-estimates and the total
range of the predictions for the six different two-flavor neutrino
scenarios that are globally consistent with all of the available
neutrino data. Figure~1 of Ref.~\cite{snoshow} shows, at $99$\% CL,
the allowed ranges of the neutrino oscillation parameters of the first
five neutrino scenarios listed in Table~\ref{tab:firstmoment}.  The
abbreviations LMA, SMA, and LOW represent three MSW solution islands
and the abbreviations ${\rm VAC_S}$ and ${\rm VAC_L}$ represent the
small-mass and large mass vacuum oscillation solutions, all for
oscillations into active neutrinos. The MSW Sterile solution has
values for the mixing angle and the square of the mass difference that
are similar to the active SMA solution (see discussion in
Ref.~\cite{snoshow}).

For a 5 MeV electron energy threshold, the predicted shifts in the
first moment, $\Delta T = \langle T\rangle - \langle T\rangle_0$, range from $-152$ keV to
$+576$ keV.  The calculational uncertainties and the measurement
uncertainties estimated from the expected behavior of SNO, $\pm 96$
keV, are considerably smaller than the total range of shifts, $711$
keV, predicted by the currently allowed set of oscillation
solutions. The shift in the first moment may be measurable if either
the SMA, ${\rm VAC_S}$, ${\rm VAC_L}$, or MSW Sterile solutions are
correct.  For the LMA and LOW solutions, the predicted shifts in the
first moment may be too small to obtain a very significant
measurement.

A measurement of the first moment with an energy threshold of 5 MeV
and a $1\sigma$ accuracy in $\langle T\rangle$ of $100$ keV or better
will significantly reduce the allowed range of neutrino oscillation
solutions.  Table~\ref{tab:firstmoment} shows that a measurement of
$\langle T\rangle$ with an energy threshold of 8 MeV will be valuable,
although it will provide a less stringent constraint than a
measurement with a lower threshold. For an 8 MeV threshold, the
currently allowed range is only $412$ keV, almost a factor of two less
than the range currently allowed for a 5 MeV threshold.

\begin{figure}[!h]
\centerline{\psfig{figure=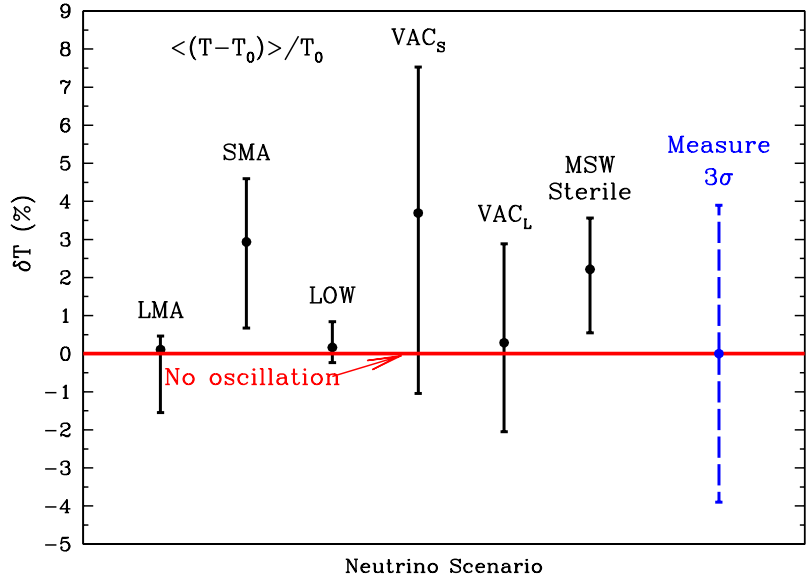,width=4.5in,angle=270}}
\tightenlines
\caption[]{\small The fractional shift in the average electron recoil
energy.  The figure shows $\delta T$ [defined in
Eq.~(\ref{eq:defndeltat})], the fractional change in the average
electron recoil energy, $\langle T\rangle$, for the six currently
allowed neutrino oscillation solutions.  The solid error bars represent the
$99$\% C.L. for the allowed regions of the six currently favored
neutrino oscillation solutions~\cite{snoshow}.  For an undistorted
$^8$B neutrino energy spectrum, the average recoil energy is denoted
by $T_0$.  The dashed error bar labeled ``Measure $3\sigma$'' represents
the uncertainty in interpreting the measurements according to the
estimates in Table~\ref{tab:uncertainties}, which include the energy
resolution, energy scale, $^8$B neutrino energy spectrum, neutrino
cross section, counting statistics, and the $hep$ flux.  The results are
calculated assuming a $5$ MeV threshold for the CC reaction.
\label{fig:deltat}}
\end{figure}

Figure~\ref{fig:deltat} shows, for a 5 MeV electron energy threshold,
the range of the fractional shift in percent of the first moment,
\begin{equation}
\delta T = \Delta T/\langle T\rangle_0, 
\label{eq:defndeltat}
\end{equation}
for all six of the
oscillation solutions. The results are compared with the 
no-oscillation solution, $\delta T = 0$.
The estimated experimental uncertainty in $\delta T$ is about $1$\%
 (see Table~\ref{tab:uncertainties}). Only the ${\rm VAC_{S}}$ solutions
predict, for about half of their currently allowed solution space, a
deviation of $\langle T\rangle$ from the no-oscillation value by more than
$3\sigma$. The MSW sterile solution predicts a shift in the first
moment that is at most $2.7\sigma$ from the no-oscillation case; this
seems like a small shift, but it is notoriously difficult to identify
measurable indications of sterile neutrinos that are different from a
reduction in the total $^8$B solar neutrino flux~\cite{sterilebilenky}.

\begin{table}
\centering
\tightenlines
\caption[]{\label{tab:secondmoment} {\bf The standard deviation, 
\boldmath$\langle \sigma^2\rangle^{1/2}$, of the
electron recoil energy spectrum.}
If there are no oscillations, the expected value of the
standard deviation of the electron recoil energy spectrum 
is $\sigma_0 = \langle\sigma^2\rangle^{1/2}_0 = 1.852$ MeV for a 5 MeV total
electron recoil energy threshold and $1.240$ MeV for an 8 MeV 
energy threshold. 
The table shows for different neutrino oscillation scenarios 
the absolute shift of the standard deviation, 
$\Delta \sigma = \langle\sigma^2\rangle^{1/2} -
\langle\sigma^2\rangle^{1/2}_0$, in keV
for both a 5 MeV total electron recoil energy threshold and an 8 MeV
energy threshold.}
\begin{tabular}{lrrrrrr} 
Scenario&${(\Delta \sigma)}_{\rm b.f.}$&${(\Delta \sigma)}_{\rm
min}$&${(\Delta \sigma)}_{\rm max}$&${(\Delta \sigma})_{\rm
b.f.}$&${(\Delta \sigma})_{\rm min}$&${(\Delta \sigma})_{\rm max}$\\
&keV&keV&keV&keV&keV&keV\\
&5 MeV&5 MeV&5 MeV&8 MeV&8 MeV&8 MeV\\
\noalign{\smallskip}
\hline
\noalign{\smallskip}
 LMA            &3  &$-19$   &9  &4  &$-13$    &11    \\
 SMA            &23 &1     &38 &23 &$3$     &34   \\
 LOW            &3  &$-3$    &13 &2  &$-3$    &10    \\
  ${\rm VAC_S}$ &70 &11    &136&44  &9     &76    \\
${\rm VAC_L}$   &127&$-29$   &199&160  &$-40$    &212    \\
 ${\rm Sterile}$&19 &2     &32 &14  &$-4$     &36  \\
\end{tabular}
\end{table}

Table~\ref{tab:secondmoment} presents the predicted shifts in the
standard deviation of the CC electron recoil energy distribution 
 (i.e., the square root of the second moment). 
The calculated no-oscillation value is 
\begin{equation}
\sigma_0 = \langle\sigma^2\rangle^{1/2}_0 = 1.852(1
\pm 0.024)\, {\rm MeV},
\label{eq:sigmanoscil}
\end{equation} 
for a 5 MeV total electron recoil energy threshold and $\sigma_0 =
1.240$ MeV for an 8 MeV threshold.
The estimated uncertainties in Eq.~(\ref{eq:sigmanoscil}) are taken from
Table~\ref{tab:uncertainties} and 
Table~III of Ref.~\cite{bkl97}. The result given in
Eq.~(\ref{eq:sigmanoscil}) is for a pure $^8$B neutrino spectrum. If a
$hep$ flux equal to the standard solar model value~\cite{bp98} is
included, the value of $\sigma_0$ is increased by $4$ keV to $1.856$
keV. For an $8$ MeV threshold,  $\sigma_0$ is increased by $6$ keV to
$1.246$ MeV by adding a nominal standard $hep$ flux.

Shifts in the standard deviation caused by neutrino oscillations will
be difficult to measure since the spread in the predicted shifts for a
$5$ MeV threshold is only from $-29$ keV to $+199$ keV, while the
estimated calculational and non-statistical measurement uncertainties
are $\pm 91$ keV. Thus the total range of the predicted shifts is less
than three standard deviations of the estimated non-statistical
uncertainties. For most of the oscillation scenarios, the shift in
$\sigma$ predicted for an 8 MeV threshold is even smaller than for a 5
MeV threshold.

It will be useful to measure the standard deviation of the recoil
energy spectrum in order to test the prediction that the observed
value will be close to the undistorted value of $\sigma_0$.

\section{The charged current rate}
\label{sec:ccrate}
In this section, we summarize the results from Ref.~\cite{snoshow} on
the expected range of predictions for the charged current (neutrino
absorption) rate [see Eq.~(\ref{eq:nuabs})].

In accordance with Eq.~(\ref{eq:defnratio}), we define the reduced
CC  neutrino-absorption  ratio [CC] by 

\begin{equation}
[{\rm CC}] ~\equiv~ \frac{{\rm (Observed~CC~Rate)}}{{\rm (Standard~CC~Rate)}}.
\label{eq:defnccratio}
\end{equation}
If the standard solar model is correct and there are no neutrino
oscillations or other non-standard physics processes, then

\begin{eqnarray}
{\rm   [CC] } & = & 1.0 \left[1.0 ~\pm 0.058^{\rm a} ~ \pm 0.019^{\rm
b} ~\pm 0.004^{\rm c} ~\pm 0.015^{\rm d} ~~~ ^{+0.023}_{-0.000}\right] \nonumber \\ 
& = &1.0\left[1.0 ~~~ ^{+0.067}_{-0.061}\right].
\label{eq:ccstandard}
\end{eqnarray}
The uncertainties in the standard solar model flux~\cite{bp98} do not affect
[CC] since we fix the absolute flux for each set of neutrino
parameters by fitting to the Super-Kamiokande total rate and recoil
energy spectrum (see discussion following Eq.~\ref{eq:definitionsfs}).

The reduced $\nu$-$e$ scattering rate 
has been measured by Super-Kamiokande~\cite{superk} to be about $0.475
\pm 0.015$ for a threshold of $6.5$ MeV and the reduced $\nu$-$e$ scattering
ratio is predicted to be approximately the same for the expected SNO
energy thresholds (see Table~\ref{tab:escratio} and Ref.~\cite{kwong96}).

The non-statistical uncertainties shown in Eq.~(\ref{eq:ccstandard})
result from (cf. Table~\ref{tab:uncertainties}): (a) the difference
between the Ying, Haxton, and Henley~\cite{yhh} and Kubodera-Nozawa
cross sections ~\cite{kn} neutrino cross sections, (b) the shape of
the $^8$B neutrino energy spectrum~\cite{b8spectrum}, (c) the energy
resolution function, (d) the absolute energy scale, and, the last
term, the uncertain $hep$ neutrino flux.  The total uncertainty in the
charged current ratio [CC] is dominated by the uncertainty in the CC
absorption cross section.

\begin{table}
\centering
\tightenlines
\caption[]{\label{tab:ccratio} {\bf The Charged Current Absorption
Ratio, [CC].}  The table presents, [CC], the ratio of the observed
neutrino absorption rate on deuterium to the standard model absorption
rate, cf. Eq.~(\ref{eq:defnccratio}).  The results are tabulated for
different neutrino oscillation scenarios and for two different
thresholds of the total electron recoil energy, 5 MeV (columns two
through four) and 8 MeV (columns five through seven). The second
(fifth) column gives the best-fit value, and the third (sixth) and
four (seventh) columns give the minimum and maximum values.}
\begin{tabular}{ddddddd} 
Scenario&[CC]&[CC]&[CC]&[CC]&[CC]&[CC]\\
&\multicolumn{1}{c}{b.f.}&max&min&\multicolumn{1}{c}{b.f.}&max&min\\
& 5 MeV& 5 MeV& 5 MeV& 8 MeV& 8 MeV& 8 MeV\\
\noalign{\smallskip}
\hline
\noalign{\smallskip}
 LMA&  0.35&0.40&0.29&0.35&0.40&0.29\\
 SMA& 0.39&0.46&0.32&0.43&0.47&0.38\\
 LOW&0.38&0.40&0.35&0.38 &0.40&0.35\\ 
 ${\rm VAC_S}$ &0.39&0.45&0.31&0.44&0.50&0.34\\
 ${\rm VAC_L}$ &0.42&0.44&0.40&0.40&0.47&0.35\\
 ${\rm Sterile}$ & 0.48&0.49&0.47&0.50&0.55&0.48
\end{tabular}
\end{table}

The most important question concerning the CC rate in SNO is the
following: Is the reduced CC rate less than the reduced
neutrino-electron scattering rate?  If the reduced CC rate is measured
to be less than the reduced $\nu$-$e$ scattering rate, then this will be
evidence for neutral currents produced by $\nu_\mu$ or $\nu_\tau$ 
which appear as a result of neutrino oscillations.  The
$3\sigma$ uncertainty is about $20$\% above the expected value of
$0.48$ (based upon the Super-Kamiokande $\nu$-$e$ scattering
measurement).  Inspecting for all six of the currently favored
oscillation solutions the range predicted for the double ratio [CC]
 (as shown in Table~\ref{tab:ccratio} or Fig.~2 of
Ref.~\cite{snoshow}), we estimate that there are very roughly equal
odds that the measured value of [CC] will lie three or more $\sigma$
below the no-oscillation value. However, for the ${\rm VAC_L}$ and MSW
Sterile solutions, the predicted range for [CC] does lie within
$3\sigma$ of the value expected on the basis of the no-oscillation
hypothesis.

\section{The neutral current rate}
\label{sec:ncrate}

We discuss in this section 
the expected range of predictions for the neutral current 
rate [see Eq.~(\ref{eq:nunc})].

If the standard solar model is correct and if there are 
either no neutrino oscillations or oscillations only into active
neutrinos, then  the reduced
NC  neutrino-absorption  rate, [NC], defined by 

\begin{equation}
[{\rm NC}] ~\equiv~ \frac{{\rm (Observed~NC~Rate)}}{{\rm (Standard~NC~Rate)}},
\label{eq:defnncratio}
\end{equation}
will satisfy

\begin{eqnarray}
{\rm   [NC] } & = & 1.0 ~\pm 0.060^{\rm a} ~ \pm 0.016^{\rm
b} ~\pm 0.02^{\rm c} ~~~^{+0.022}_{-0.000}~~~^{+0.18}_{-0.16}\nonumber \\ 
& = &1.0 \pm 0.07~ ~ ~^{+0.18}_{-0.16}.
\label{eq:ncstandard}
\end{eqnarray}
The non-statistical uncertainties shown in Eq.~(\ref{eq:ncstandard})
result from (cf. Table~\ref{tab:uncertainties}): (a) the difference
between the Ying, Haxton, and Henley~\cite{yhh} and Kubodera-Nozawa
cross sections ~\cite{kn} neutrino cross sections, (b) the shape of
the $^8$B neutrino energy spectrum~\cite{b8spectrum}, and (c) the
uncertainty in the neutral current detection efficiency. 
The next to last 
term in Eq.~(\ref{eq:ncstandard}) represents the uncertainty in the
$hep$ neutrino flux.

The last term in Eq.~(\ref{eq:ncstandard}) represents the uncertainty
in the BP98 standard $^8$B flux~\cite{bp98}.  In our method, this
uncertainty only appears if there are no neutrino oscillations or
other new physics. If there are neutrino oscillations, we determine
the ratio, $f(^8B)$.  of the best-fit neutrino flux to the standard
model flux as described in Sec.~\ref{sec:introduction} following
Eq.~(\ref{eq:defnratio}) and in Sec.~\ref{sec:shapespectrum} following
Eq.~(\ref{eq:defrofe}) and Eq.~(\ref{eq:definitionsfs}).
  
The total uncertainty in determining experimentally the neutral
current ratio [NC] is dominated by the uncertainty in the NC
absorption cross section and the total uncertainty in interpreting the
neutral current measurement is dominated by the uncertainty in the
predicted solar model flux.

If one assumes that only oscillations into active
neutrinos occur, then it will be possible to use the neutral current
measurement as a test of the solar model calculations.
The cross section uncertainty for the NC reaction is about $33$\%
 ($38$\% ) of the upper (lower) estimated uncertainty in the solar
model flux. 

If the SMA Sterile neutrino
solution~\cite{bks98,snoshow,four} is correct, then for the
global solutions acceptable at $99$\% C.L.,

\begin{equation}
[{\rm NC}] = 0.465 \pm 0.01   ~.
\label{eq:ncsterile}
\end{equation}
Unfortunately, this result is within about $3\sigma$ of the result
expected if there are oscillations into active neutrinos, when one
includes the solar model uncertainty shown in
Eq.~(\ref{eq:ncstandard}).

\section{The neutrino-electron scattering rate}
\label{sec:esc}

In this section, we present the predictions for the
electron-scattering rate, Eq.~(\ref{eq:nuesc}), in SNO. The SNO event
rate for this process is expected to be small, $\sim$ 10\% of the CC
rate. Despite the relatively unfavorable statistical uncertainties,
the measurement of the neutrino-electron scattering rate in SNO will
be important for two reasons. First, the measurement of the electron
scattering rate by SNO will provide an independent confirmation of the
results from the Kamiokande and Super-Kamiokande experiments. Second,
the neutrino-electron scattering rate can be combined with other
quantities measured in SNO so as to decrease the systematic
uncertainties and to help isolate the preferred neutrino oscillation
parameters.

If the best-estimate standard solar model neutrino flux~\cite{bp98} is
correct and there are no new particle physics effects, then the
reduced neutrino-electron scattering rate, [ES], will be measured to
be
\begin{equation}
[{\rm ES}] ~\equiv~ \frac{{\rm (Observed~\nu-e~Rate)}} {{\rm (Standard~\nu-e~Rate)}}   ~=~1.0 \pm 0.02,
\label{eq:defnescratio}
\end{equation}
where the non-statistical uncertainties are taken from
 Table~\ref{tab:uncertainties}.  For a $6.5$ MeV threshold and the
 experimental parameters of the Super-Kamiokande
 detector~\cite{superk}, ${\rm [ES]_{SK}} = 0.475 \pm 0.015$.

The uncertainties summarized in Eq.~(\ref{eq:defnescratio}) include
 all the uncertainties given in Table~\ref{tab:uncertainties} except
 for statistical errors. 
The uncertainties in the standard solar model flux~\cite{bp98} do not affect
[ES] since we fix the absolute flux for each set of neutrino
parameters by fitting to the Super-Kamiokande total rate and recoil
energy spectrum (see discussion following Eq.~\ref{eq:definitionsfs}).
The  dominant non-statistical uncertainties are from the value of the
 $hep$ flux and the shape of the $^8$B neutrino energy spectrum. For
 the first five or ten years of the SNO operation, the overall
 dominant error in the determination of [ES] is expected to be
 statistical: $\geq 5$\% after the first 5000 CC events.

Table~\ref{tab:escratio} gives, for two different energy thresholds,
the values of the reduced neutrino-electron scattering rate, [ES],
that are predicted by the currently favored oscillation
scenarios~\cite{snoshow} . Not surprisingly, the values of [ES]
cluster around the ratio measured by the Super-Kamiokande experiment
$[{\rm ES}]_{\rm SK} = 0.475 \pm 0.015$\cite{superk}. The global
constraints imposed by the different experiments result in some cases
in the spread of the currently favored predictions being less than or
of the order the total spread in the Super-Kamiokande rate measurement.
\begin{table}
\centering
\tightenlines
\caption[]{\label{tab:escratio} {\bf The Neutrino-Electron Scattering
Ratio, [ES].}  The table presents, [ES], the ratio of the observed
neutrino-electron scattering rate to the standard model rate,
cf. Eq.~(\ref{eq:defnescratio}).  The results are tabulated for
different neutrino oscillation scenarios and for two different
thresholds of the total electron recoil energy, 5 MeV (columns two
through four) and 8 MeV (columns five through seven). The second
(fifth) column gives the best-fit value, and the third (sixth) and
four (seventh) columns give the minimum and maximum values.}
\begin{tabular}{ddddddd} 
Scenario&[ES]&[ES]&[ES]&[ES]&[ES]&[ES]\\
&\multicolumn{1}{c}{b.f.}&max&min&\multicolumn{1}{c}{b.f.}&max&min\\
& 5 MeV& 5 MeV& 5 MeV& 8 MeV& 8 MeV& 8 MeV\\
\noalign{\smallskip}
\hline
\noalign{\smallskip}
 LMA&  0.48&0.49&0.47&0.47&0.48&0.45\\
 SMA& 0.47&0.48&0.46&0.50&0.51&0.48\\
 LOW&0.48&0.48&0.47&0.48 &0.49&0.47\\ 
 ${\rm VAC_S}$ &0.48&0.52&0.46&0.51&0.54&0.46\\
 ${\rm VAC_L}$ &0.49&0.50&0.47&0.47&0.50&0.45\\
 ${\rm Sterile}$ & 0.46&0.48&0.45&0.51&0.53&0.48
\end{tabular}
\end{table}

\begin{figure}[!t]
\centerline{\psfig{figure=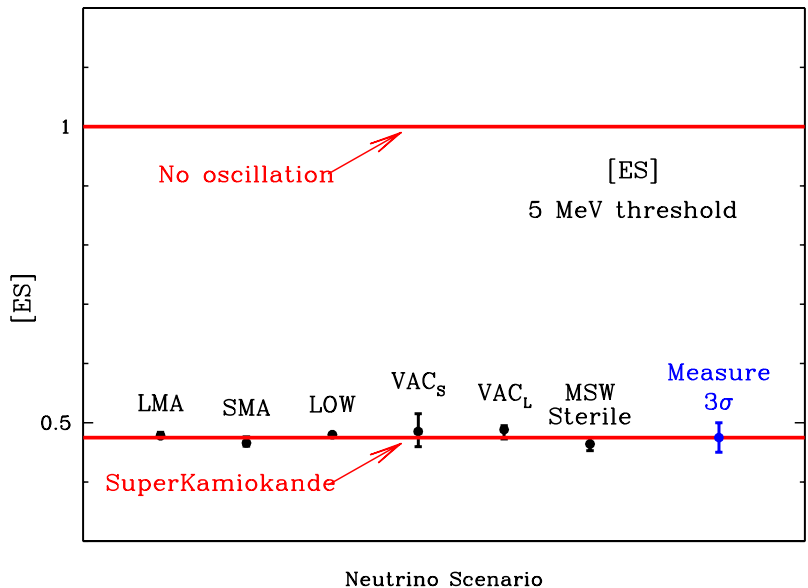,width=4.5in,angle=270}}
\tightenlines
\caption[]{\small The reduced neutrino-electron scattering rate, [ES].
The figure shows the reduced neutrino-electron scattering rate
[cf. Eq.~(\ref{eq:defnescratio})] from the standard model value of
$1.0$ (no oscillations).  The Super-Kamiokande result (see
Ref.~\cite{superk}) is also shown. The solid error bars represent the
$99$\% C.L. for the allowed regions of the six currently favored
neutrino oscillation solutions~\cite{snoshow}.  The dashed error bar
labeled ``Measure $3\sigma$'' represents the uncertainty in
interpreting the measurements according to the estimates in
Table~\ref{tab:uncertainties}, which include the energy resolution,
energy scale, $^8$B neutrino energy spectrum, neutrino cross section,
counting statistics, and the $hep$ flux.  The oscillation predictions
are calculated assuming a $5$ MeV threshold for the total electron
recoil energy.
\label{fig:esc}}
\end{figure}

Figure~\ref{fig:esc} compares the oscillation predictions for
[ES]$_{\rm SNO}$ versus [ES]$_{\rm SuperK}$ and the no-oscillation
solution. We only show the predictions for a $5$ MeV threshold for the
total electron energy since the results are similar for an $8$ MeV
threshold (see Table~\ref{tab:escratio}).  The solid error bars shown
in Fig.~\ref{fig:esc} reflect the range at $99$\% C.L. of the globally
allowed solutions that are fit to all the available neutrino
data~\cite{snoshow}.

\section{The neutral current to charged current double ratio}
\label{sec:ncovercc}

In this section, we present  predictions for the ratio of neutral
current events (NC) to charged current events (CC) in SNO. The most
convenient form in which to discuss this quantity is obtained by dividing
the observed ratio by the ratio computed assumed the correctness of
the standard electroweak model (SM). This double ratio is defined by
the relation~\cite{howwell}

\begin{equation}
{\rm  {[NC]} \over { [CC]} } = 
{ 
{\left({\rm (NC)_{Obs}/(NC)_{SM} }\right) } \over 
{\left({\rm (CC)_{Obs}/(CC)_{SM} }\right) } 
}.
\label{eq:defnncovercc}
\end{equation}
The ratio ${\rm {[NC]}/ { [CC]} }$ is equal to unity if nothing
happens to the neutrinos after they are produced in the center of the
sun (no oscillations occur). Also, ${\rm {[NC]} / { [CC]} }$ is
independent of all solar model considerations provided that  only
one neutrino source, $^8$B, contributes significantly to the measured
rates. Finally, the calculational uncertainties due to the interaction
cross sections and to the shape of the $^8$B neutrino energy spectrum
are greatly reduced by forming the double ratio (see
Table~\ref{tab:uncertainties}).

Table~\ref{tab:doubleratio} presents the calculated range of the
double ratios for the oscillation solutions that are currently allowed
at $99$\% CL~\cite{snoshow}.  The table gives the best-fit values 
for  ${\rm {[NC]} / { [CC]} }$ as
well as the maximum and minimum allowed double ratios for a total
electron energy threshold for the CC reaction of $5$ MeV and
separately for a CC threshold of $8$ MeV.

\begin{table}
\centering
\tightenlines
\caption[]{\label{tab:doubleratio} {\bf Neutral Current to Charged Current
Double Ratio.}  The table presents the double ratio, ${\rm [NC]/[CC]}$, that
is defined by Eq.~(\ref{eq:defnncovercc}).  The results are tabulated
for different neutrino oscillation scenarios and for two different
thresholds of the total electron recoil energy used in computing the
CC ratio, 5 MeV (columns two through four) and 8 MeV (columns five
through seven).}
\begin{tabular}{ddddddd} 
Scenario&[NC]/[CC]&[NC]/[CC]&[NC]/[CC]&[NC]/[CC]&[NC]/[CC]&[NC]/[CC]\\
&\multicolumn{1}{c}{b.f.}&max&min&\multicolumn{1}{c}{b.f.}&max&min\\
& 5 MeV& 5 MeV& 5 MeV& 8 MeV& 8 MeV& 8 MeV\\
\noalign{\smallskip}
\hline
\noalign{\smallskip}
 LMA&  3.37&5.15&2.27&3.36&5.13&2.30\\
 SMA& 2.53&4.11&1.24&2.28&3.50&1.21\\
 LOW& 2.71&3.39&2.30&2.69&3.35&2.26\\ 
 ${\rm VAC_S}$ & 2.67&4.63&1.37&2.33&3.73&1.27\\
 ${\rm VAC_L}$ & 1.90& 2.15&1.53&2.01&2.45&1.53\\
 ${\rm Sterile}$ & 0.96&0.99&0.94&0.88&0.97&0.82
\end{tabular}
\end{table}

\begin{figure} [!h]
\centerline{\psfig{figure=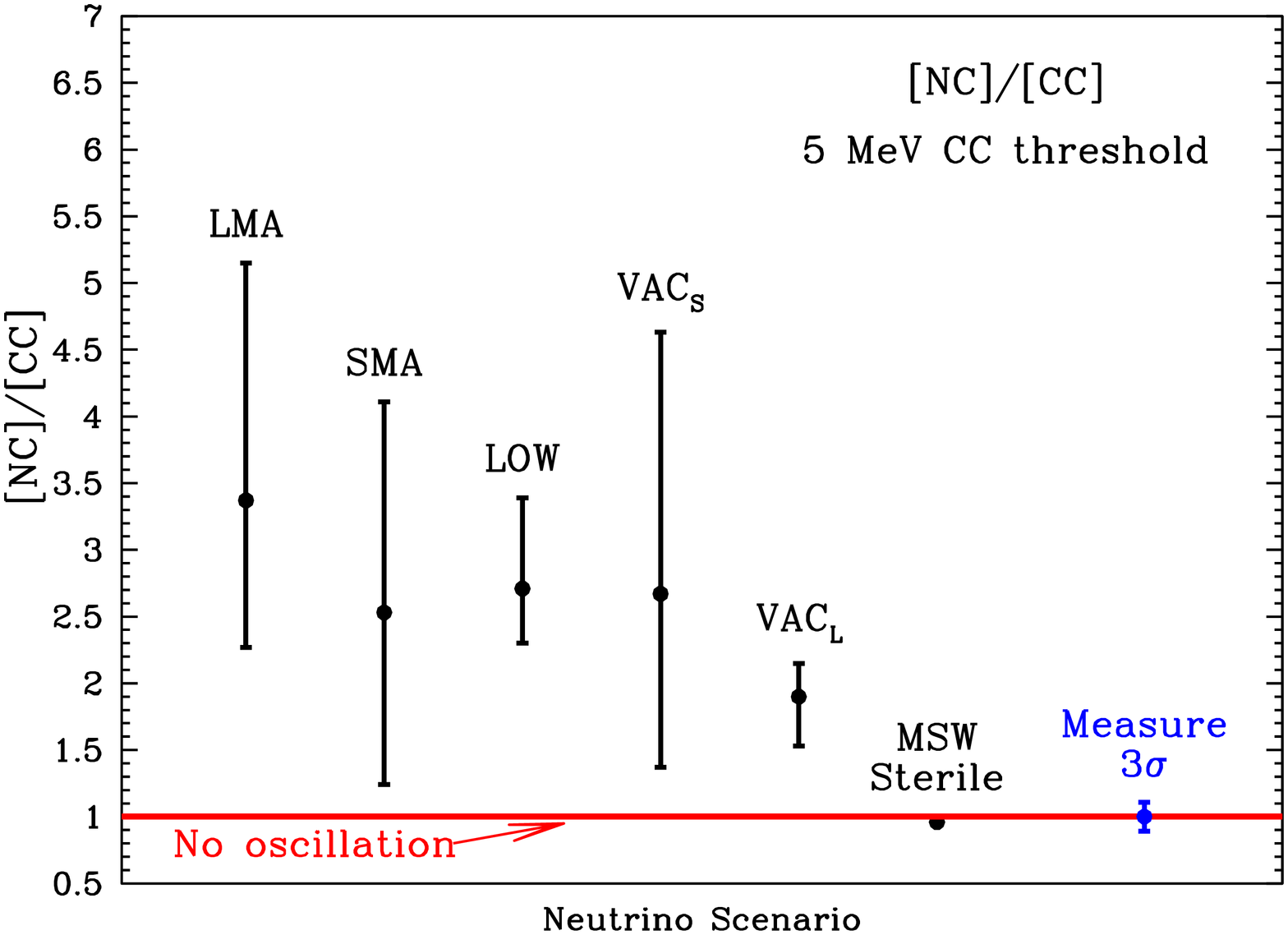,width=4.5in,angle=270}}
\centerline{\psfig{figure=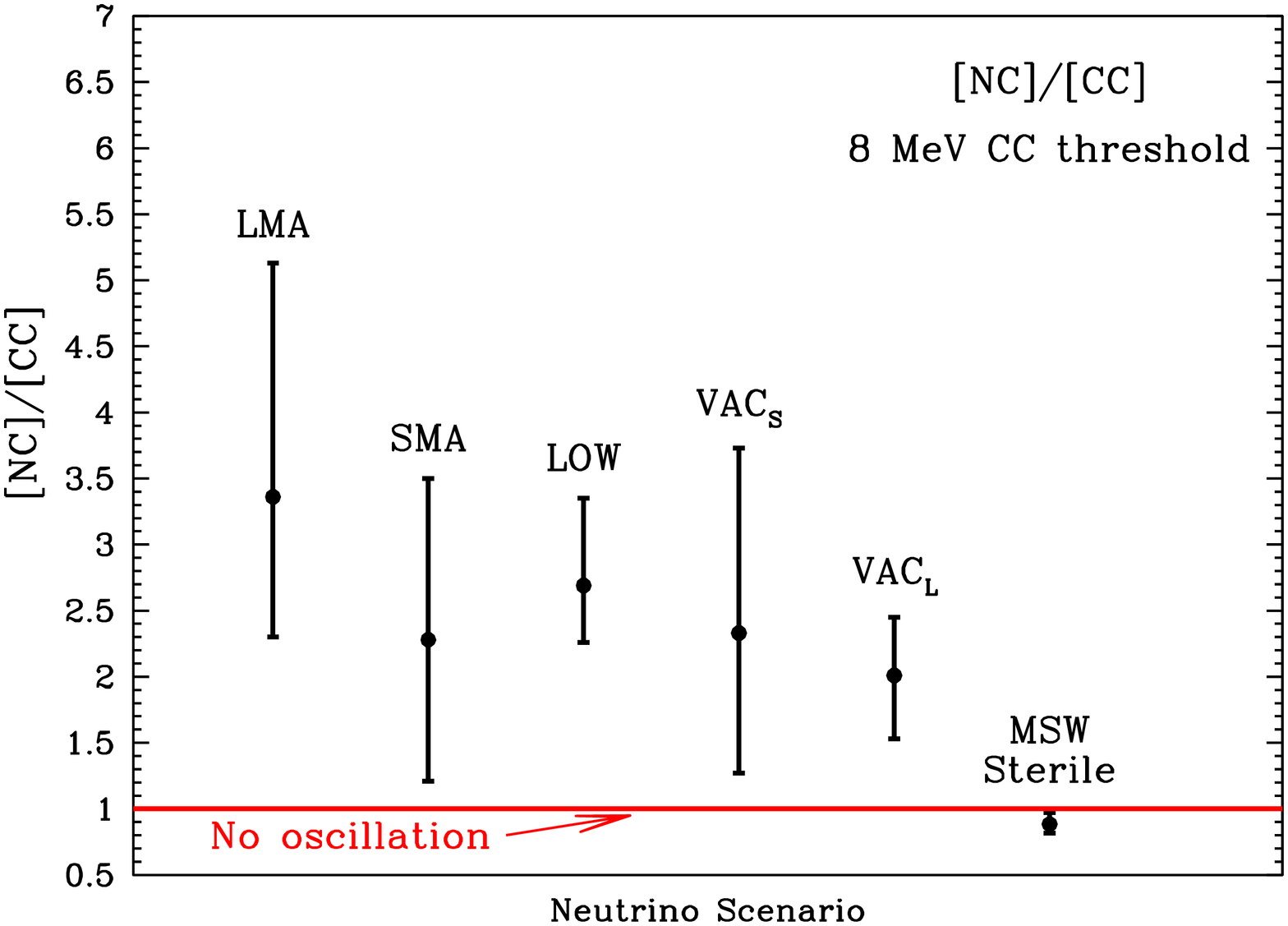,width=4.5in,angle=270}}
\tightenlines
\caption[]{\small The neutral current to charged current double ratio.
The double ratio, [NC]/[CC] is defined by Eq.~(\ref{eq:defnncovercc}).
The standard model value for [NC]/[CC] is $1.0$.
Figure~\ref{fig:double}a shows, for a 5 MeV threshold for the CC
measurement, the predicted double ratio of Neutral Current to Charged
Current for different neutrino scenarios.  Figure~\ref{fig:double}b
shows the same ratio but for an 8 MeV CC threshold.  The solid error bars
shown represent the $99$\% C.L. for the allowed regions of the six
currently favored neutrino oscillation solutions~\cite{snoshow}. 
The dashed error bar labeled ``Measure $3\sigma$'' represents
the uncertainty in interpreting the measurements according to the
estimates in Table~\ref{tab:uncertainties}, which include the energy
resolution, energy scale, $^8$B neutrino energy spectrum, neutrino
cross section, counting statistics, and the $hep$ flux.
\label{fig:double}}
\end{figure}

Figure~\ref{fig:double} compares the predicted values of [NC]/[CC]
with the no-oscillation value of 1.0. The results are shown for a 5
MeV CC threshold and for an 8 MeV CC threshold. 
The estimated [see Eq.~(\ref{eq:ncccstandard})] non-statistical
errors are smaller than the black dots indicating the
best-fit points in Fig.~\ref{fig:double}.

The best-fit values for the double ratio for oscillations into active
neutrinos range between $1.9$ and $3.4$ ($2.0$ and $3.4$) for a 5 MeV
 (8 MeV) CC threshold.  The maximum predicted values for ${\rm {[NC]} /
{ [CC]} }$ exceed $5.1$.  For active neutrino oscillations, the
minimum values for the double ratio are achieved by the SMA and the
${\rm VAC_L}$ solutions; they are $1.2$ and $1.5$, respectively.

The sterile neutrino solutions predict a double ratio in a band that
is separate from all the active oscillation solutions, namely, $0.9$
to $1.0$.  The physical reason that the double ratio for sterile
neutrinos is less than $1.0$ is that in the SMA solution (for active
or sterile neutrinos) the probability that a solar $\nu_e$ survives as
a $\nu_e$ decreases with energy (see, e.g., Fig.~ 9 of
Ref.~\cite{bks98}). In the sterile neutrino case, if the $\nu_e$
oscillates to another state it does not interact. Since the NC
threshold is $2.2$ MeV and the observational threshold for CC events
is likely to be $5$ MeV or above, the smaller survival probability at
low energies more strongly affects the average NC rate than the
average CC rate.

The standard model value for ${\rm {[NC]} / { [CC]} }$ is

\begin{eqnarray}
{\rm  {[NC]} \over { [CC]} } & = & 1.0 ~\pm 0.004^{\rm a} ~ \pm 0.003^{\rm
b} ~\pm 0.004^{\rm c} ~\pm 0.015^{\rm d} ~\pm 0.02^{\rm e} \nonumber \\ 
& = &1.0 \pm
0.026. 
\label{eq:ncccstandard}
\end{eqnarray}
The uncertainties, all non-statistical, shown in
Eq.~(\ref{eq:ncccstandard}) result from: (a) the difference between
the Ying, Haxton, and Henley~\cite{yhh} and Kubodera-Nozawa cross
sections ~\cite{kn}, (b) the shape of the $^8$B neutrino energy
spectrum~\cite{b8spectrum}, (c) the energy resolution function, (d)
the absolute energy scale, and (e) the NC detection efficiency.
Comparable, but small, contributions are made by the cross section
uncertainties, the uncertainties in the shape of the neutrino energy
spectrum, and the uncertainty in the energy resolution function.  The
absolute energy scale and the NC detection efficiency are expected to
contribute even less to the errors.

One of the principal uncertainties in interpreting the electron recoil
energy spectrum is the poorly known value for the flux of the
extremely rare $hep$ neutrinos~\cite{bkhep,frere}. The uncertainty in
the $hep$ flux can also affect the otherwise robust measurement of
${\rm{[NC]}/{[CC]}}$ (see Table~\ref{tab:uncertainties}).  We have
recalculated the value of ${\rm{[NC]}/{[CC]}}$ for a $hep$ flux that
is $20$ times larger than the nominal standard model
flux~\cite{bp98}. We find [cf. Eq.~(\ref{eq:hepuncertain}) for the
calculational prescription] 

\begin{equation}
{\rm{[NC]} \over {[CC]}}~=~1.0 
-\epsilon_{hep} \left({\phi(hep)} \over
{20 \, \phi(hep,~{\rm BP98}) }\right) ,
\label{eq:nccchep}
\end{equation}
where $\epsilon_{hep} = 0.0005$ for a 5 MeV threshold on the CC events
and $0.017$ for an 8 MeV CC threshold.  For an 5 MeV CC threshold,
there is an accidental cancellation of the contributions to the
neutral current ratio, [NC], and to the charged current ratio, [CC],
so that the net value of $\epsilon_{hep}({\rm 5~MeV})$ is very
small. But, for an 8 MeV threshold, the $hep$ flux causes an
uncertainty, $1.7$\%, that is larger than the combined contribution
from all the other known uncertainties except possibly counting
statistics [cf. Eq.~(\ref{eq:ncccstandard}) and Eq.~(\ref{eq:nccchep})
and Table~\ref{tab:uncertainties}]. Of course, the gain at lower
energies due to the reduction in the uncertainty from the $hep$
neutrinos may be more than offset by the increased uncertainty due to
background events.  Fortunately, SNO is expected to be able to measure
or to place strong limits on the $hep$ flux within the first full year
of operation~\cite{snoshow}.

We have not included the statistical uncertainties in the
calculational error budget of Eq.~(\ref{eq:ncccstandard}). It seems
likely that statistical errors will dominate over calculation errors,
at least in the first several years of operation of SNO (see
Table~\ref{tab:uncertainties}). The CC rate may be in the range of
3000 to 4000 events per year.  The NC event rate in the detector will
be about a factor of two smaller and the NC detection rate will be
further decreased by the NC detection efficiency that may be of order
$50$\%.  Thus statistical errors in the NC rate, the uncertainty
($\sim 2$\%, see Table~\ref{tab:uncertainties}) in the NC detection
rate, and the uncertainties in the $hep$ flux, will probably be the
limiting factors in determining the accuracy of the experimental
measurement of [NC]/[CC].

The small calculational error [see Eq.~(\ref{eq:ncccstandard})] for
${\rm {[NC]} / { [CC]} }$, combined with the relatively large
differences between the no-oscillation and the oscillation values for
active neutrinos (Table~\ref{tab:doubleratio}), makes the double
ratio an ideal `smoking-gun' indicator of oscillations.

\section{The Electron-scattering to CC Double Ratio}
\label{sec:esctocc}

In this section, we present results for the double ratio of neutrino-electron
scattering to CC events. This ratio is defined, by analogy
with the NC to CC double ratio [see Eq.~(\ref{eq:defnncovercc})], by the
expression

\begin{equation}
{\rm  {[ES]} \over { [CC]} } = 
{ 
{\left({\rm (ES)_{Obs}/(ES)_{SM} }\right) } \over 
{\left({\rm (CC)_{Obs}/(CC)_{SM} }\right) } 
}.
\label{eq:defnescovercc}
\end{equation}

The double ratio ${\rm{[ES]}/{ [CC]}}$ has some of the same
advantages as the double ratio ${\rm{[NC]}/{ [CC]}}$, namely,
independence of solar model considerations and partial cancellation of
uncertainties. In fact, the ${\rm{[ES]}/{ [CC]}}$ double ratio has
the additional advantage that the same detection process is used for
the recoil electrons from both the scattering and the CC
reactions. For the ${\rm{[NC]}/{ [CC]}}$ double ratio, different
techniques are used to determine the two rates and this increases the
systematic measurement uncertainty in the ratio.

\begin{table}
\centering
\tightenlines
\caption[]{\label{tab:escdoubleratio} {\bf Neutrino-Electron Scattering to
Charged Current Double Ratio.}  The table presents the double ratio,
${\rm [ES]/[CC]}$, that is defined by Eq.~(\ref{eq:defnescovercc}).
The results are tabulated for different neutrino oscillation scenarios
and for two different thresholds of the total electron recoil energy
for both the scattering and the CC reactions, 5 MeV (columns two
through four) and 8 MeV (columns five through seven).}
\begin{tabular}{ddddddd} 
Scenario&[ES/CC]&[ES/CC]&[ES/CC]&[ES/CC]&[ES/CC]&[ES/CC]\\
&\multicolumn{1}{c}{b.f.}&max&min&\multicolumn{1}{c}{b.f.}&max&min\\
& 5 MeV& 5 MeV& 5 MeV& 8 MeV& 8 MeV& 8 MeV\\
\noalign{\smallskip}
\hline
\noalign{\smallskip}
 LMA&  1.37&1.65&1.20&1.36&1.63&1.19\\
 SMA& 1.20&1.43&1.03&1.29&1.57&1.05\\
 LOW&1.27 &1.37&1.20&1.26& 1.37&1.20\\ 
 ${\rm VAC_S}$ &1.25  &1.58&1.03&1.16&1.38&1.02\\
 ${\rm VAC_L}$ &1.15  &1.20&1.08&1.18&1.29&1.05\\
 ${\rm Sterile}$ & 0.96&0.99&0.92&1.06&1.08&1.02
\end{tabular}
\end{table}

The standard model value for ${\rm{[ES]}/{ [CC]}}$ is

\begin{eqnarray}
{\rm  {[ES]} \over { [CC]} } & = & 1.0 ~\pm 0.06^{\rm a} ~ \pm 0.006^{\rm
b} ~\pm 0.003^{\rm c} ~\pm 0.01^{\rm d} ~\nonumber \\ 
& = &1.0 \pm
0.06, 
\label{eq:escccstandard}
\end{eqnarray}
where the non-statistical uncertainties shown in
Eq.~(\ref{eq:escccstandard}) result from: (a) the difference between
the Ying, Haxton, and Henley~\cite{yhh} and Kubodera-Nozawa cross
sections ~\cite{kn} neutrino cross sections, (b) the shape of the
$^8$B neutrino energy spectrum~\cite{b8spectrum}, (c) the energy
resolution function, and (d) the absolute energy scale.  The upper
limit $hep$ uncertainty is small, $0.007$ (see
Table~\ref{tab:uncertainties}).

The
uncertainty in the double ratio [ES]/[CC] is dominated by the
uncertainty in the CC absorption cross section and by statistical
errors ($\sim 4.9$\% after 5000 CC events, see
Table~\ref{tab:uncertainties}) that are not included in
Eq.~(\ref{eq:escccstandard}).

Table~\ref{tab:escdoubleratio} presents the calculated range of
${\rm{[ES]}/{ [CC]}}$ for the oscillation solutions that are
currently allowed at $99$\% CL~\cite{snoshow}.  The table gives the
best-fit values for ${\rm {[ES]}/{ [CC]}}$ as well as the maximum and
minimum allowed double ratios for a total electron energy threshold
 (for both reactions) of either $5$ MeV or $8$ MeV.  The range of
ratios predicted by oscillations into active neutrinos is $1.03$ to
$1.65$, much smaller than the range ($1.24$ to $5.1$) predicted for the
${\rm {[NC]}/{ [CC]}}$ double ratio.

\begin{figure}
\centerline{\psfig{figure=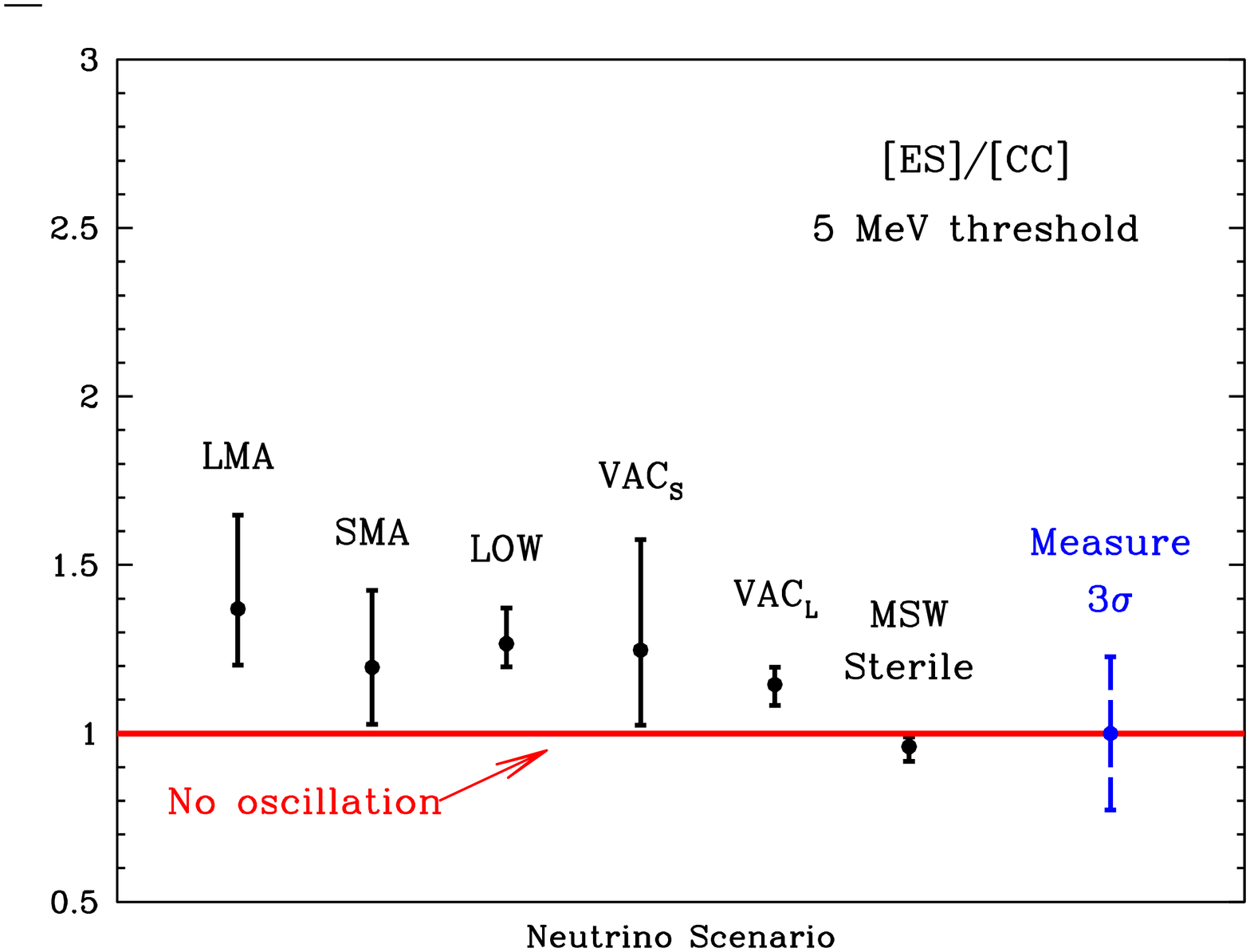,width=4.5in,angle=270}}
\centerline{\psfig{figure=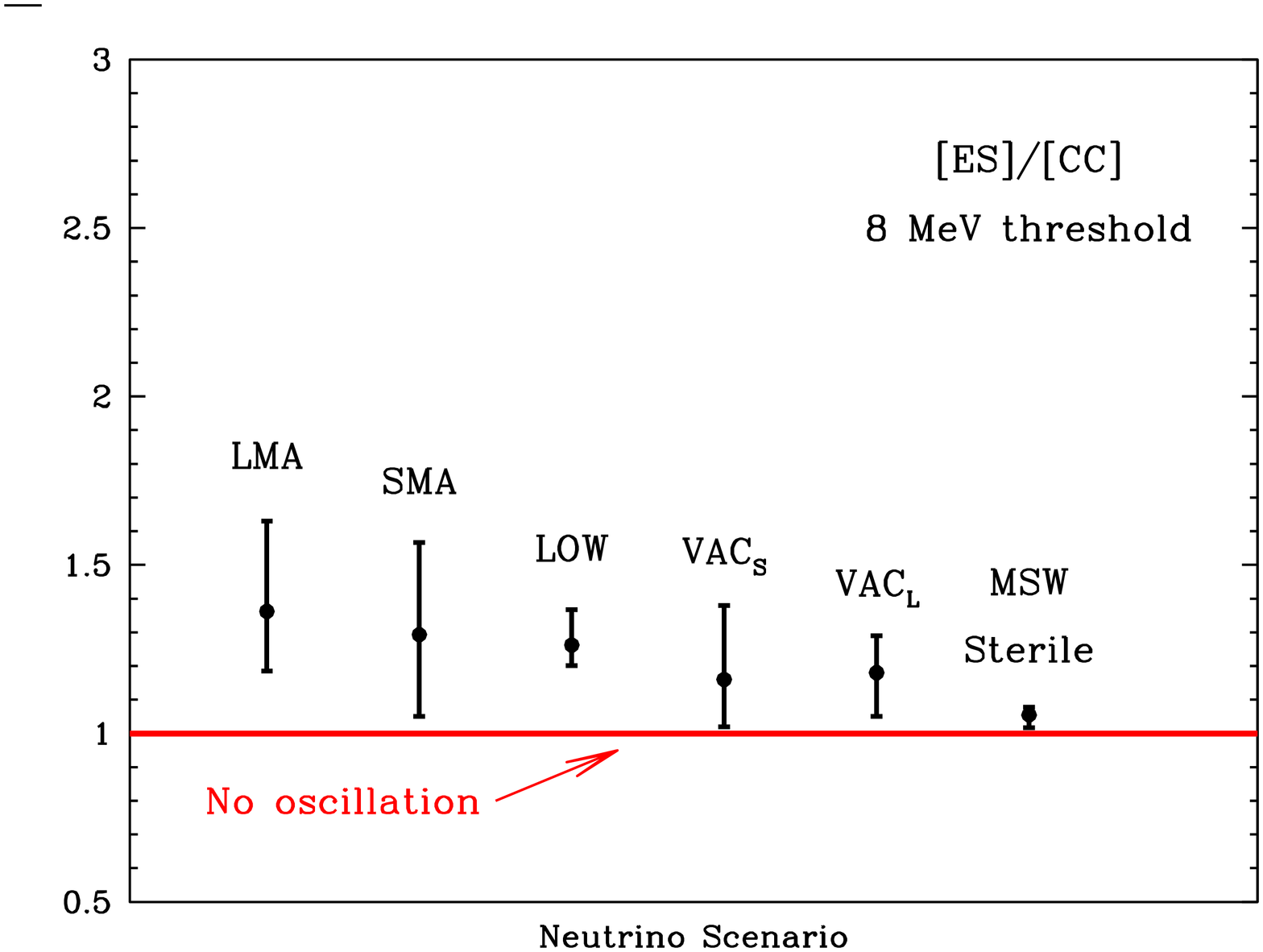,width=4.5in,angle=270}}
\tightenlines
\caption[]{\small The neutrino-electron scattering to charged current
double ratio.  The double ratio, [ES]/[CC] is defined by
Eq.~(\ref{eq:defnescovercc}).  Figure~\ref{fig:escdouble}a shows, for
a 5 MeV electron recoil energy threshold, the predicted double ratio
of Neutrino-Electron Scattering to Charged Current for different
neutrino scenarios.  The solid error bars represent the $99$\%
C.L. for the allowed regions of the six currently favored neutrino
oscillation solutions~\cite{snoshow}.  The dashed error bar labeled
``Measure $3\sigma$'' represents the uncertainty in interpreting the
measurements according to the estimates in
Table~\ref{tab:uncertainties}, which include the energy resolution,
energy scale, $^8$B neutrino energy spectrum, neutrino cross section,
counting statistics, and the $hep$ flux.  Figure~\ref{fig:double}b
shows the same ratio but for an 8 MeV threshold.
Figure~\ref{fig:escdouble}, which has a vertical scale extending from
$0.5$ to $3.0$, should be compared with Fig.~\ref{fig:double}, which
has a vertical scale extending from $0.5$ to $7.0$.  The difference
from the no-oscillation solution is much greater for the [NC]/[CC]
double ratio than it is for the [ES]/[CC] double ratio.
\label{fig:escdouble}}
\end{figure}

Figure~\ref{fig:escdouble} shows the values of [ES]/[CC] predicted by
the different oscillation solutions.  Comparing
Fig.~\ref{fig:escdouble} and Fig.~\ref{fig:double}, one can see that
the neutral current to charged current ratio is a more sensitive
diagnostic of neutrino oscillations than is the electron scattering to
charged current ratio. The difference from the no-oscillation solution
is much greater for the [NC]/[CC] double ratio than it is for the
[ES]/[CC] double ratio. In addition, there are expected to be many
more detected NC events than neutrino-electron scattering events.
Also, the cross section uncertainties largely cancel out of the ratio
${\rm{[NC]}/{ [CC]}}$, whereas the uncertainty in the CC cross section
is an important limitation in interpreting the ratio [ES]/[CC].

\section{The day-night effect}
\label{sec:daynight}

We discuss in this section the difference between the event rate
observed at night and the event rate observed during the day.  For MSW
solutions, the interactions with matter of the earth can change the
flavor content of the solar neutrino beam and cause the nighttime and
daytime rates to differ.  This effect has been discussed and evaluated
by many different authors, including those listed in
Ref.~\cite{daynight}.

We concentrate here on the difference, $A_{\rm N-D}$, between the
nighttime and the daytime rates, averaged over one year. The formal
definition of $A_{\rm N-D}$ is
\begin{equation}
A_{\rm N-D} ~=~2{\rm  {[Night - Day]} \over {[Night + Day]}}.
\label{eq:daynightdefn}
\end{equation}
In what follows, we shall use $A_{\rm N-D}$ to refer to the charged
current reaction. When we want to consider the quantity defined by
Eq.~(\ref{eq:daynightdefn}) for the neutral current, we shall write 
$A_{\rm N-D}({\rm NC})$.

 We begin by discussing in Sec.~\ref{subsec:nooscdaynight} the
apparent day-night effect that arises solely from the eccentricity of
the earth's orbit and the inclination of the earth's axis (the existence of
seasons), and then we discuss in Sec.~\ref{subsec:ccdaynight}
the day-night effect for the CC reaction and in
Sec.~\ref{subsec:ncdaynight} the day-night effect for the NC reaction
due to oscillations.

\subsection{The No-Oscillation day-night effect}
\label{subsec:nooscdaynight}

In the absence of neutrino oscillations, there is a geometrical
day-night effect that we have not seen discussed in previous
publications. This No-Oscillation (NO) effect is caused by the
ellipticity of the earth's orbit and by the fact that, in the northern
hemisphere, nights are longer (days are shorter) in winter when the
earth is closer to the sun. Thus the average over the year of the
nighttime rate will be larger than the annual average of the daytime
rate for all detectors located in the northern hemisphere.

We find that the No-Oscillation (NO) day-night effect is 

\begin{eqnarray}
A^{\rm No}_{\rm N-D} &~=~&0.0094~~({\rm SNO}),
\label{eq:snonooscdaynight} \\
A^{\rm No}_{\rm N-D} &~=~~&0.0066~~({\rm SK}), 
\label{eq:sknooscdaynight} \\
A^{\rm No}_{\rm N-D} &~=~&0.0082~~({\rm Gran~Sasso}),
 \label{eq:grannooscdaynight} \\
A^{\rm No}_{\rm N-D} &~=~&0.0088~~({\rm Homestake}), 
 \label{eq:homenooscdaynight} 
\end{eqnarray}
for the locations of the SNO, Super-Kamiokande, Gran Sasso, and
Homestake detectors.

The No-Oscillation effect is  purely geometrical; it 
is independent of neutrino energy and independent of neutrino flavor.
The magnitude of the NO effect is the same for the CC, ES, and NC reactions.
The numerical results given in Eq.~(\ref{eq:sknooscdaynight})--
Eq.~(\ref{eq:homenooscdaynight}) can also be obtained from the
following easily-derived relation, which makes clear the seasonal
aspect of the NO effect:

\begin{equation}
A^{\rm No~ Osc.}_{\rm N-D} \approx 2 \epsilon { {\left(t_{\rm max} -
t_{\rm min} \right)} \over {24} },
\label{eq:analyticnodaynight}
\end{equation} 
where $\epsilon = 0.0167$ is the eccentricity of the
earth's orbit and $t_{\rm max}$ and $t_{\rm min}$ are, respectively,
the length of the longest and the shortest nights in the year at the
location of the detector.

In what follows, we remove the No-Oscillation day-night effect before
presenting the predictions of an additional day-night effect that is
due to neutrino oscillations.  More precisely, we calculate the
day-night effect assuming that the neutrino flux from the sun is
constant throughout the year.  The effects that we discuss in
Sec.~\ref{subsec:ccdaynight} and in Sec.~\ref{subsec:ncdaynight} are
due to neutrino mixing.

Experimental results can easily be analyzed so as to remove the NO
day-night effect. All that is required is to multiply the number of
events in each time bin ($\Delta t \ll $ 1 year) by the ratio
$\left[r(t)/(1~{\rm A.U.})\right]^2$, where $r(t)$ is the average
earth-sun distance in that time bin and 1~A.U. is the annual
average earth-sun distance.  This is the procedure adopted by the
Super-Kamiokande collaboration~\cite{superk}.

Even after these corrections, there is a residual day-night effect for
vacuum oscillations. In this case, the day-night effect is due to the
variation of the survival probability as a function of the distance
between the earth and the sun and the fact that in the northern
hemisphere the longest nights occur when the earth is closest to the
sun.  We are not aware of any previous discussions of the day-night
effect for vacuum oscillations. For MSW oscillations, the day-night
effect is caused by neutrino flavor changes during propagation in the
earth.

\subsection{The CC day-night effect}
\label{subsec:ccdaynight}

\begin{figure}[!ht]
\centerline{\psfig{figure=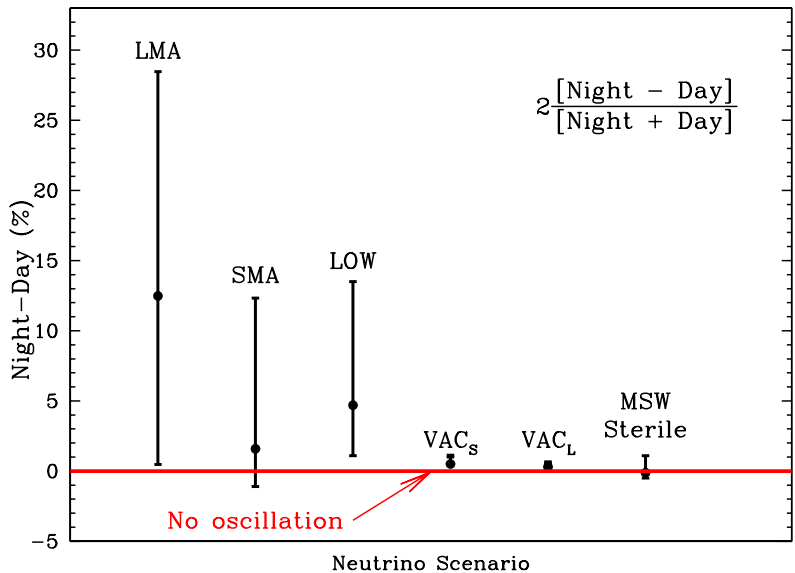,width=4.5in,angle=270}}
\tightenlines
\caption[]{\small The percentage difference between the night and the
day CC rates.  The figure shows for different neutrino scenarios the
percentage difference between the predicted CC rate in SNO at night
and the CC rate in the day [see Eq.~(\ref{eq:daynightdefn})]. The
error bars represent the $99$\% C.L. for the allowed regions of the
six currently favored neutrino oscillation solutions~\cite{snoshow}.
The results were calculated with a CC threshold of 5 MeV for the total
electron recoil energy.
\label{fig:daynight}}
\end{figure}

Table~\ref{tab:daynight} and Fig.~\ref{fig:daynight} present the range
of predicted percentage differences between the average rate at night
and the average rate during the day [i.e., $100 \times A_{\rm N-D}$ of
Eq.~(\ref{eq:daynightdefn})].  The calculated predictions are given
for a $5$ MeV and an $8$ MeV CC electron recoil energy threshold.

For vacuum oscillations, the day-night effect is due to the dependence
of the survival probability upon the earth-sun distance. 
The predicted day-night effect for vacuum oscillations is small in all
the cases shown in Table~\ref{tab:daynight} and in Fig.~\ref{fig:daynight}.

For most of the MSW oscillation solutions, the predicted
day-night differences are only of order a few percent. However, for
the LMA solution, the predicted difference can reach as high as $29$\%
for a $5$ MeV threshold ($32$\% for an $8$ MeV threshold).  There are
also rather large differences, in excess of $10$\%, that are possible
for the SMA and LOW solutions.

At first glance, one might think that such large day-night differences
will be easy to measure. In fact, there are important systematic
uncertainties that have to be taken into account in making sure that
the relative sensitivities to the day and the night rates are properly
evaluated~\cite{superk}. Even the purely statistical uncertainties are
very significant because the day-night difference, $A_{\rm N-D}$, is
the difference between two comparably sized large numbers. Thus the
fractional statistical uncertainty after accumulating a large number,
$N$, of counts at night (and a roughly equal number during the day) is

\begin{equation}
{\sigma (A_{\rm N-D}) \over A_{\rm N-D}} 
~\simeq~  \left({1 \over A_{\rm N-D} }\right) {\sqrt{2 \over
N}} .
\label{eq:ncccstatistical}
\end{equation}
The fact that $A_{\rm N-D}$ can be a small number makes a multi-sigma
statistical measurement of the day-night effect difficult.  The
careful analysis of the day-night effect by the Super-Kamiokande
collaboration~\cite{superk} has demonstrated the practical difficulty
of a precision measurement of $A_{\rm N-D}$.  Using more than $800$
effective days of operation of the SuperK detector with total night
time counts of $N \sim 5900$ ($\sim 11,200$ total events) the precision
obtained by the Super-Kamiokande collaboration is $A_{\rm SK}~=~0.065
\pm 0.03$, i.e., $\Delta A /A \sim 0.5$.  To accumulate with SNO an
equivalent number of CC events ($\sim 11,000$ total events) may
require of order three years or longer of operation.

\begin{table}[!t] 
\centering
\tightenlines
\caption[]{\label{tab:daynight} {\bf The Predicted Night-Day
Difference (in \%) for the CC Rate.} The table gives the percentage
difference in the Night-Day CC rates, $A_{\rm N-D}$, defined by
Eq.~(\ref{eq:daynightdefn}).  For a $5$ MeV ($8$ MeV) threshold energy
for recoil electrons, the second (fifth) column gives the
best-fit value, and the third (sixth) and fourth (seventh) columns give
the minimum and maximum values, allowed at 99\% CL for different
neutrino oscillation solutions (cf. Fig. 1 of Ref.~\cite{snoshow}).}
\begin{tabular}{ldddddd} 
Scenario&\multicolumn{1}{c}{b.f.}&min&max&\multicolumn{1}{c}{b.f.}&min&max\\
\noalign{\smallskip}
\hline
\noalign{\smallskip}
 LMA&           12.4    &$+$0.5     &28.5 & 14.1 &0.7 &31.6 \\
 SMA&           1.6     &$-$1.1     &12.3 &1.2 & $-$1.3& 10.9\\
 LOW&            4.7    &$+$1.1     &13.5 &4.0 &0.8 &11.8 \\ 
 ${\rm VAC_S}$&  0.5     &$-$0.1     & 1.0  & 0.7 & 0.2 & 1.2\\
 ${\rm VAC_L}$&  0.3     &$-$0.1     & 0.5 & 1.0 &$-$0.1 & 1.5 \\
 MSW, Sterile&  $-$0.1   &$-$0.4     &1.1 & $-$0.2 &$-$0.7 & 1.1\\
\end{tabular}
\end{table}

Why is the predicted effect in SNO (see also Ref.~\cite{bari}) so much
larger than for Super-Kamiokande? The reason is that for
neutrino-electron scattering the day-night effect is
decreased relative to the pure CC mode by the contribution of the
neutral currents. For the LMA solution, one can derive a simple
quantitative relation between the CC day-night effect, $A^{\rm CC}$,
and the ES day-night effect, $A^{\rm ES}$. Let the nighttime rate be
proportional to $P_{\rm N} + r(1 - P_{\rm N})$, where $P_{\rm N}$ is
the average (over energy) survival probability during the night and
$r$ is the average ratio of $\nu_\mu-e$ to $\nu_e-e$ scattering cross
sections.  Writing a similar expression for the daytime rate, it is
easy to show that

\begin{equation}
A^{\rm CC}_{\rm N-D} ~=~ A^{\rm ES}_{\rm N-D} \left[
1 +  { {r} \over {(1 - r)P} }  \right],
\label{eq:ascaling}
\end{equation}
where $P$ is the average of the day and the night survival
probabilities. Since $r \approx 0.16$ and $P \approx 0.3$ for the
best-fit solution, we see that the term in brackets in
Eq.~(\ref{eq:ascaling}) is about $1.6$. For the LMA solution, the
best-fit predicted value for $A^{\rm SK}$ is $8$\% for a $5$ MeV
recoil energy threshold, which corresponds to about $13$\% for SNO, in
good agreement with the value of $12.5$\% given in
Table~\ref{tab:daynight}. There are small corrections to
Eq.(~\ref{eq:ascaling}) due to the energy dependence of the various
neutrino quantities (and the different locations on the earth of the
SNO and the Super-Kamiokande detectors).

\subsection{The NC day-night effect}
\label{subsec:ncdaynight}

There is no day-night effect in the NC for oscillations into active
neutrinos. All active neutrinos are recorded with equal probability by
the neutral current detectors. However, for oscillations into sterile
neutrinos there can be a day-night effect since the daughter (sterile)
neutrinos are not detectable.  Thus a day-night effect in the NC would
be a `smoking gun' indication of sterile neutrino oscillations.

For the region that is allowed at
$99$\% CL by a global fit of the MSW sterile neutrino solution 
to all the available neutrino data~\cite{snoshow}, we find a NC MSW
sterile neutrino day-night effect of 

\begin{equation}
A({\rm NC,~MSW~Sterile}) ~=~-0.001^{+0.006}_{-0.002} .
\label{eq:dnsterile}
\end{equation}
Although the predicted  effect is small, it is important in principle since
there are very few ways that sterile neutrino oscillations can be
identified uniquely~\cite{sterilebilenky}.

The neutral current day-night effect for solar neutrinos was first
pointed out in Ref.~\cite{kpq}.  Here we have calculated accurately 
the predicted range of the day-night asymmetry given the latest solar
neutrino data, solar model, and a realistic model of the earth.

\section{Seasonal effects}
\label{sec:seasonal}

We discuss in this section the seasonal dependences that are predicted
by the currently favored neutrino oscillation solutions.  We define a
Winter-Summer Asymmetry by analogue with the Night-Day difference. Thus

\begin{equation}
A_{\rm W - S} ~=~2{\rm  {[Winter - Summer]} \over {[Winter + Summer]}}.
\label{eq:wintersummerdefn}
\end{equation}

The earth's motion around the sun causes a seasonal dependence that
can be calculated and is 
\begin{equation}
A_{\rm W-S,~ orbital} = 0.064~~({\rm 45~day~averages})
\label{eq:orbitalshort}
\end{equation}
 for a 45 day Winter interval centered around December 21 and a $45$ day
Summer interval centered around June 21. The average length of the
winter (summer) night during this $45$ day period is $15.4$ ($8.5$)
hours.  The amplitude is reduced if the entire year is divided into
two parts, with the winter average being taken as $182$ days centered
on December 21 and with the average length of the winter (summer)
night being $14.1$ ($9.7$) hours.  In this case, the asymmetry is
reduced by a factor of $1.5$ from the $45$-day value.  Thus
\begin{equation}
A_{\rm W-S,~ orbital} = 0.042~~({\rm 182~day~averages}).
\label{eq:orbitallong}
\end{equation}

In what follows, we have removed the seasonal dependence due to the
orbital motion from the quoted values of the seasonal dependence due
to neutrino oscillation effects.

\begin{figure}[!ht]
\centerline{\psfig{figure=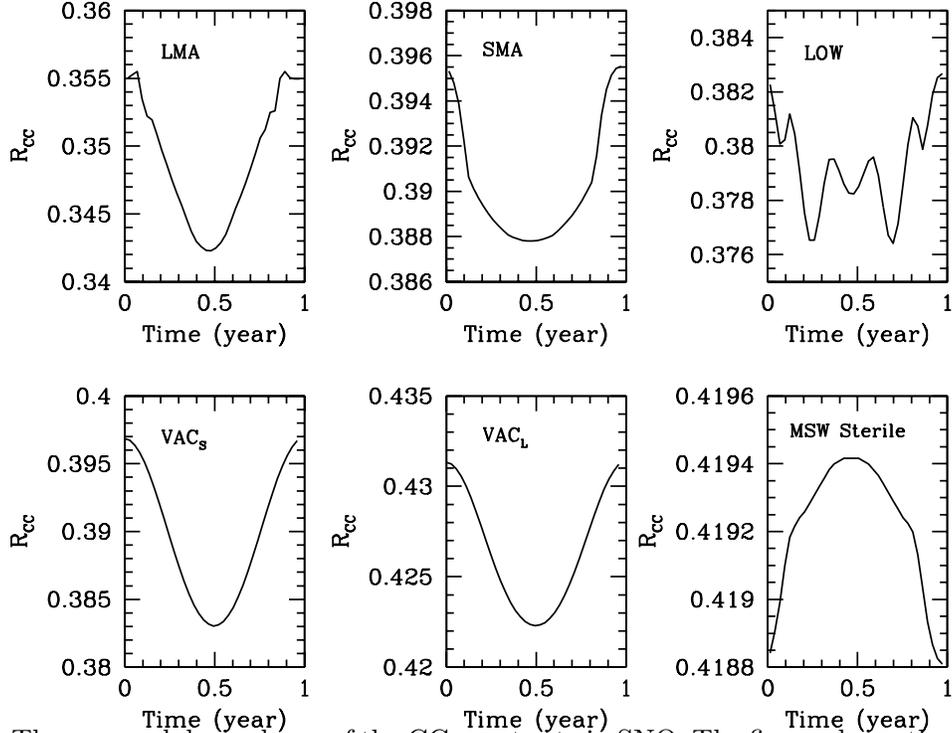,width=5in,angle=0}}
\tightenlines
\caption[]{\small The seasonal dependence of the CC event rate in SNO.
The figure shows the dependence on the time of the year of the
predicted CC event rate in SNO for the six best-fit neutrino
oscillation solutions described in Table~\ref{tab:bestfit}. 
The time labeled zero on the horizontal axis corresponds to January 1.
\label{fig:seasonal}}
\end{figure}

Figure~\ref{fig:seasonal} shows the predicted dependence upon the day
of the year of the CC event rate, $R_{\rm cc}$, in SNO for each of the
currently favored best-fit oscillation solutions.  The vertical scales
are different, reflecting the fact that the predicted seasonal
variations are, e.g., relatively large for the best-fit LMA and ${\rm
VAC_L}$ solutions, but are tiny for the MSW sterile solution.  The
annual average of the events rates shown in Fig.~\ref{fig:seasonal}
yields the numbers shown in the second column of
Table~\ref{tab:ccratio}. The alert reader may notice that the phases
of the variations in the two panels referring to vacuum oscillations
are shifted by about two weeks with respect to the four panels that
refer to MSW oscillations. This shift results from the fact that the
earth and the sun are closest (relevant for vacuum oscillations) on
January 4 and the day with the longest night is December 21 (relevant
for MSW oscillations).

\begin{figure}[!ht]
\centerline{\psfig{figure=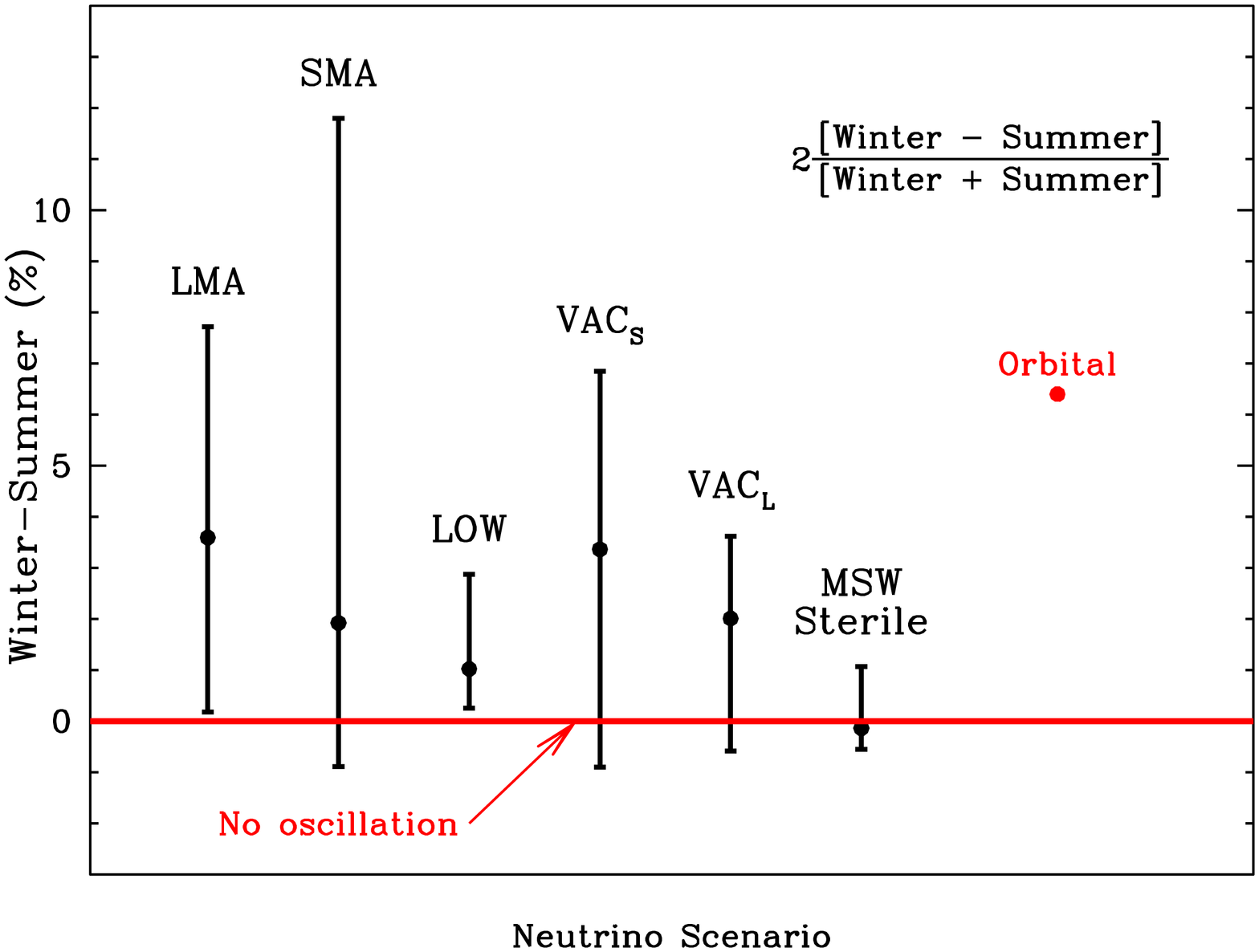,width=5in,angle=270}}
\tightenlines
\caption[]{\small The difference between the winter and the summer
rates.  The figure shows for different neutrino scenarios the
percentage difference between the predicted CC rate in SNO for a 45
day interval during winter and the CC rate for a 45 day interval in
the summer [see Eq.~(\ref{eq:wintersummerdefn})].  The point labeled
`Orbital' represents the 45 day winter-summer difference due to the
earth's motion around the sun; the amplitude of this orbital motion
has been removed from the neutrino oscillation points.  The error bars
represent the $99$\% C.L. for the allowed regions of the six currently
favored neutrino oscillation solutions~\cite{snoshow}.  The results
were calculated with a CC threshold of 5 MeV for the total electron
recoil energy.
\label{fig:wintersummer}}
\end{figure}

Table~\ref{tab:wintersummer} and Fig.~\ref{fig:wintersummer} show
the calculated percentage amplitudes for the 45 day winter-summer
difference due to oscillations, $A_{\rm W - S}$. 
In all cases, the best-fit oscillation solutions predict a
winter-summer difference due to neutrino properties that is less than
the orbital effects given in Eq.~(\ref{eq:orbitalshort}) and
Eq.~(\ref{eq:orbitallong}). Only rather extreme cases give amplitudes
of $A_{\rm W - S}$ due to oscillations that are as large as the
orbital amplitude, which will itself require a number of years to establish
definitively~\cite{sno}.

\begin{table}[!t] 
\centering
\begin{minipage}{4in}
\tightenlines
\caption[]{\label{tab:wintersummer} {\bf The Winter-Summer Predicted
Difference in the CC Rate.} The table gives the percentage difference,
$A_{\rm W-S}$, in the Winter-Summer CC rates, defined by
Eq.~(\ref{eq:wintersummerdefn}).  The results on the first row for
each oscillation solutions have been computed for $45$ days in winter
and $45$ days in the summer; the results on the second row are average
over $182$ days of winter and $182$ days of summer.  The second column
gives the best-fit value, and the third and four columns give the
minimum and maximum values, allowed at 99\% CL for different neutrino
oscillation solutions (cf. Fig. 1 of Ref.~\cite{snoshow}). The
winter-summer asymmetry due to the earth's motion around the sun,
$A_{\rm W-S,~orbital} = 6.4\% (4.2\%)$ for a $45$ day ($182$ day)
average, has been removed from the values given here.}
\begin{tabular}{lddd} 
Scenario&\multicolumn{1}{c}{b.f.}&min&max\\
Interval&45 d&45 d&45 d\\
        &182 d&182 d&182 d\\
\noalign{\smallskip}
\hline
\noalign{\smallskip}
 LMA&           3.6      &0.2       &7.7 \\
    &            2.4      &0.1       &4.7 \\
 SMA&           1.9      &$-$0.9      &11.8 \\
    &            1.1      &$-$0.7      &7.8  \\ 
LOW&           1.0      &0.25      &2.9  \\ 
    &          0.6      &0.15      &1.7  \\
 ${\rm VAC_S}$&  3.4     &$-$0.9     & 6.9  \\
              &  2.2     &$-$0.6     & 4.5  \\
 ${\rm VAC_L}$&  2.0     &$-$0.6     & 3.6  \\
              &  1.3     &$-$0.4     & 2.4  \\
 MSW, Sterile&  $-$0.1     &$-$0.55     & 1.1  \\
             &  $-$0.1     &$-$0.3      &0.55  \\
\end{tabular}
\end{minipage}
\end{table}

Table~\ref{tab:wintersummer} also gives the predicted values of
$A_{\rm W - S}$ for a longer average, $182$ of winter and $182$ of
summer. For this case the statistical error will be reduced by about a
factor of two, but the size of the effect is typically reduced by a
factor of order $1.5$ to $1.7$.    

For the LMA solution, we showed in Ref.~\cite{lma} that to a good
approximation $A_{\rm W-S}$ and $A_{\rm N-D}$ are related by the equation

\begin{equation}
A_{\rm W-S} = A_{\rm N-D} \left[\frac{t_W - t_S}{24~{\rm h} }\right],
\label{eq:correlation}
\end{equation}
where $t_W$ and $t_S$ are the average lengths of the nights during
the selected winter and summer periods, respectively.  For the $45$ day
 ($182$ day) intervals we are discussing here, the length at SNO of the
winter night is $15.43$ hours ($14.10$ hours) and the length of the
summer night is $8.49$ hours ($9.72$ hours).  The term in brackets in
Eq.~(\ref{eq:correlation}) is $1.6$ times larger for the $45$ day
period (longer nights) than for the $182$ day period. This accounts
well for the ratios of $A_{\rm W-S}$ for the $45$ day and the $182$
periods that are given in Table~\ref{tab:wintersummer} .
Eq.~(\ref{eq:correlation}) also produces well the individual values of
$A_{\rm W-S}$ for the LMA solution. Using the best-fit value of $ A_{\rm N-D}
= 12.48$\%  (from Table~\ref{tab:daynight}) and $(t_W - t_S)/(24~{\rm
h}) = 0.29$, we estimate $A_{\rm W-S} = 3.6$\% for the $45$ day average,
in good agreement with the result given in
Table~\ref{tab:wintersummer}\footnote{For the LOW solution,
Eq.~(\ref{eq:correlation}) also gives a crude estimate of $A_{\rm W-S}$,
accurate to $\sim 40$\%.  The value of $\delta m^2$ is smaller for the
LOW than for the LMA solutions and therefore the typical oscillation
length in matter is larger. The averaging of the oscillation effects
required for the validity of Eq.~(\ref{eq:correlation}) (see
Ref.~\cite{lma}) is not complete for the LOW solution. }.

From the size of the predicted effects shown in
 Table~\ref{tab:wintersummer} and Fig.~\ref{fig:wintersummer}, we
 conclude that it will require many years of SNO operation  to measure 
an accurate value of $A_{\rm W - S}$ if the currently favored
 oscillation solutions are correct.

\section{Smoking gun vs. smoking gun}
\label{sec:smokinggun}

What do we gain by combining the measurements of different smoking gun
quantities? Once SNO has begun to report results for a variety of
different quantities and an accurate Monte Carlo of the experimental
facility exists, then it will be possible to analyze simultaneously a
variety of different measurements using a global analysis method like
Maximum Likelihood. In the meantime, we begin an initial illustrative
exploration by analyzing pairs of SNO measurements.

We show in this section how comparisons of the measurements of
different smoking-gun quantities versus each other can enhance the
deviation of a single measurement from the no-oscillation expectation
and also shrink the globally-allowed range of the oscillation
parameters.  We concentrate on the most powerful pairwise combinations
of variables.  We do not illustrate all possible combinations,
omitting some examples (like day-night effect versus first moment of
the CC energy spectrum) that turn out to be less useful when
examined quantitatively.

We begin by displaying and discussing in
Sec.~\ref{subsec:ncccdratios} the predicted oscillation regions in
planes defined by the double ratio [NC]/[CC] versus either 1) the
day-night effect, $A$; 2) the first moment, $\langle T\rangle$, of the CC recoil
energy spectrum; and 3) the neutrino-electron scattering reduced rate,
[ES].  The double ratios involving the neutral current discriminate
sharply between oscillation and no-oscillation scenarios and also
reduce the range of acceptable oscillation parameters.  In
Sec.~\ref{subsec:escccvst} and Sec.~\ref{subsec:ccratevst}, we
discuss the location of the favored oscillation solutions in the
[ES]/[CC] versus $\langle T\rangle$ plane and in the plane of the CC rate, $R_{\rm
CC}$ versus the first moment, $\langle T\rangle$.

For each plane defined by two SNO parameters and for each of the six
neutrino oscillation solutions, we plot error bars that represent
separately the $99$\% C.L. acceptable range of the neutrino parameters
in the global fits to all the currently available solar neutrino
data~\cite{snoshow}. The $1\sigma$ experimental uncertainties are
summarized in Table~\ref{tab:uncertainties}.  The statistical
uncertainties are computed assuming $5000$ CC events, $1219$ NC
events, and $458$ ES events. We assume that the $hep$ uncertainties
are symmetric and equal to the upper limit uncertainty, which slightly
increases the error contours. For the no-oscillation case, only the
experimental measurements are correlated.  When neutrino oscillations
occur, the predicted values for different parameters are also
correlated.  We include here only the correlations of the
uncertainties for the no-oscillation case; we do not include the
correlated contours for the six different predicted oscillation
solutions.  A full calculation that includes the theoretical
correlations between the different measured parameters and also
includes asymmetric $hep$ uncertainties should be carried out in the
future, but this study is beyond the scope of the present paper.

The correlations between different estimated experimental
uncertainties cause the no-oscillation error ellipses to be tilted in
Fig.~\ref{fig:ncccvsa}--Fig.~\ref{fig:rccvst}. For purposes of
illustration, we have assumed that the error correlations are as
estimated in Ref.~\cite{bl}.  As we shall see from
Fig.~\ref{fig:ncccvsa}--Fig.~\ref{fig:rccvst}, the tilt of the error
ellipses can significantly influence the total statistical
C.L. assigned to a given set of results and therefore accurate
determinations of the error correlations for the SNO experiment will be
important.

\subsection{[NC]/[CC] double ratio versus other smoking guns}
\label{subsec:ncccdratios}

\subsubsection{[NC]/[CC] versus the day-night effect}
\label{subsubsec:ncccvsdaynight}

\begin{figure}[!h]
\centerline{\psfig{figure=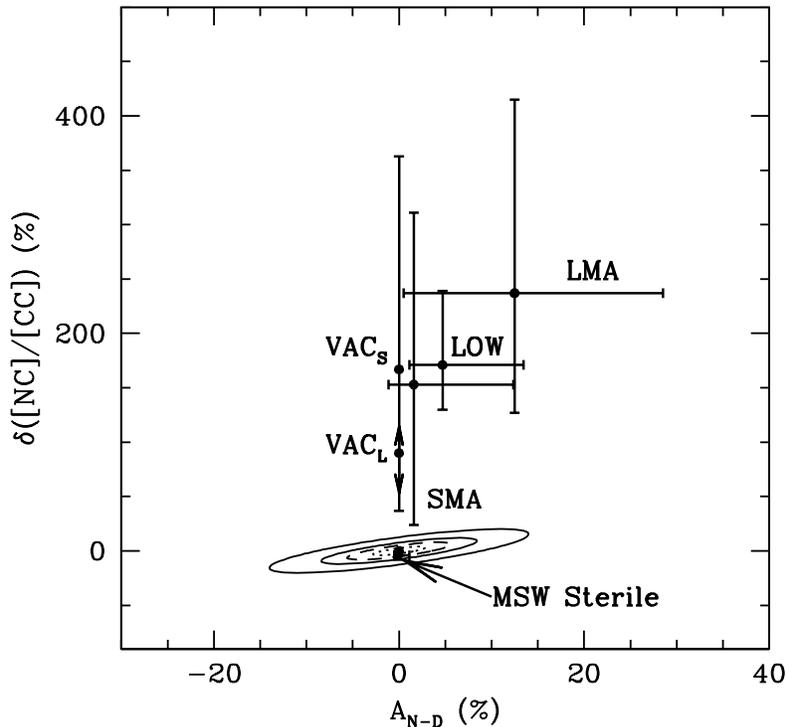,width=4in}} \tightenlines
\caption[]{\small The [NC]/[CC] double ratio versus the day-night
asymmetry. The figure shows the fractional difference in percent of
the neutral current versus charged current double ratio, [NC]/[CC]
 (Eq.~\ref{eq:defnncovercc}), from the no-oscillation value of $1.0$,
versus the day-night difference, $A_{\rm N-D}$
 (Eq.~\ref{eq:daynightdefn})in percent. The standard model value of
$A_{\rm N-D}$ is  $0.0$.
For the six currently preferred oscillation solutions, the error bars
represent the $99$\% C.L. regions for acceptable fits to all the
available neutrino data. Estimated error contours for SNO are shown
at the $1\sigma,2\sigma,3\sigma$, and $5\sigma$ levels
 relative to the no-oscillation solution which lies at $(0,0)$.
\label{fig:ncccvsa}}
\end{figure}

Figure~\ref{fig:ncccvsa} shows the values predicted by the different
oscillation solutions in the plane of the ${\rm [NC]/[CC]}$ double
ratio and the day-night asymmetry, $A_{\rm N-D}$.  Specifically, we
plot the fractional shift in percent of the [NC]/[CC] double ratio
from the standard model value of $0$\% on the vertical axis and the
predicted value in percent of the day-night asymmetry $A_{\rm N-D}$
 (standard model value of $0$\%) on the horizontal plane. Each of the
currently allowed solutions, with the exception of the MSW Sterile
solution, predicts points in the $\delta {\rm [NC]/[CC]}$--$A_{\rm N-D}$
plane that are more than $5\sigma$ separated from the standard model
solution (which is located at $0\%,0\%$). Moreover, the vacuum
solutions are separated from the MSW solutions by amounts that exceed
the expected measuring errors in $A_{\rm N-D}$ and [NC]/[CC]. It will,
however, be more difficult to distinguish between different MSW
solutions in the $\delta {\rm [NC]/[CC]}$-$A_{\rm N-D}$ plane,
although some measured values would point to a unique solution.  For
example, a large positive value of $A_{\rm N-D}$ ($\geq 20$\%)
combined with a large value of [NC]/[CC] ($\geq 2.3$) would favor the
LMA solution.  The allowed region for the MSW Sterile solution is all
contained within the ellipse corresponding to the estimated $3\sigma$
experimental uncertainty.

\subsubsection{[NC]/[CC] versus $\langle T\rangle$}
\label{subsubsec:versust}

\begin{figure}[h]
\centerline{\psfig{figure=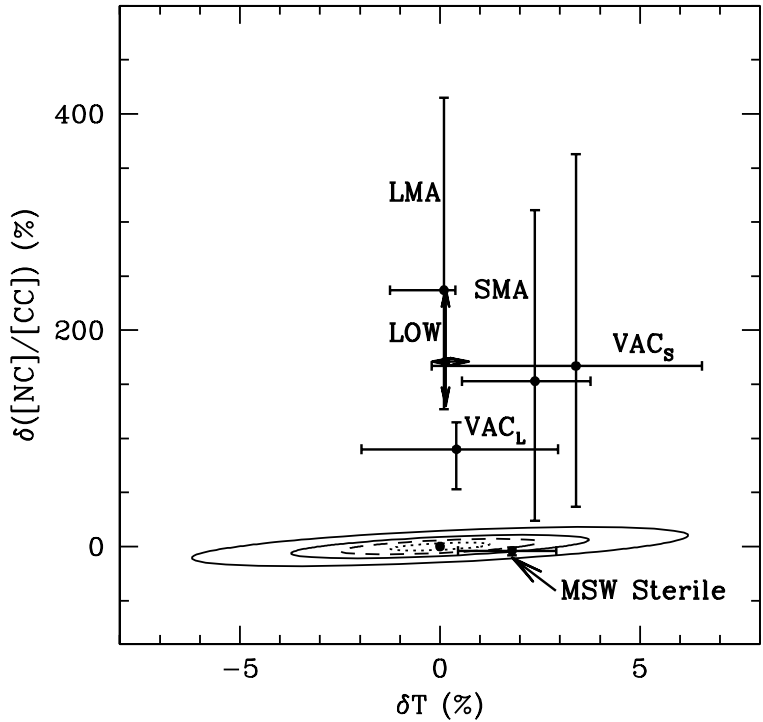,width=4in}} \tightenlines
\caption[]{\small The [NC]/[CC] double ratio versus the average recoil
energy.  The figure shows the fractional difference in percent of the
neutral current versus charged current double ratio, [NC]/[CC]
 [Eq.~(\ref{eq:defnncovercc})], from the no-oscillation value of $1.0$,
versus the fractional difference in percent of the average electron
recoil energy, $\langle T\rangle$ [Eq.~(\ref{eq:defndeltat})] , from 
the no-oscillation value [Eq.~(\ref{eq:t0})]. Contours at
$1\sigma,2\sigma,3\sigma$, and $5\sigma$ are shown relative to 
the no-oscillation solution at $(0,0)$.
\label{fig:ncccvst}}
\end{figure}

Figure~\ref{fig:ncccvst} shows the predictions of the different
oscillation solutions in the $\delta{\rm [NC]/[CC]}$ versus $\delta T$
plane.  All of the currently favored oscillation solutions, with the
exception of the MSW sterile solution, predict locations in the
$\delta{\rm [NC]/[CC]}$ versus $\delta T$ plane that are separated
by more than $5\sigma$ from the standard model solution, which lies at
 ($0.0,0.0$).  However, the discrimination is almost entirely due to
the [NC]/[CC] double ratio. The value of $\delta T$ only adds a large
discrimination for the extreme ${\rm VAC_S}$ solution. The predicted values
for the MSW Sterile solution extend out to $2.9$\%, which because of
the correlation of the experimental errors (which gives rise to the
tilt of the error ellipses in Fig.~\ref{fig:ncccvst}), can correspond
to deviations as large as $3\sigma$ from the no-oscillation solution.

\subsubsection{[NC]/[CC] versus [ES]}
\label{subsubsec:versusesc}

The most likely value for SNO to observe for the neutrino-electron
scattering ratio [ES] is close to the Super-Kamiokande~\cite{superk}
value of ${\rm [ES}_{SK} ] = 0.475$ (see for example
Table~\ref{tab:escratio} or Fig.~\ref{fig:esc}).  It is therefore
convenient to define the quantity $\delta ({\rm [ES]})$ as follows:
\begin{equation}
\delta \left({\rm ES} \right)
 ~\equiv~ \frac{{\rm [ES]_{OBS} -0.475}}{{0.475}}. 
\label{eq:defnescminussk}
\end{equation}
We have used parentheses rather than squared brackets in defining
$\delta \left({\rm ES} \right)$ because the shift in [ES] is
measured relative to $0.475$ rather than $1.0$ .

\begin{figure}{!t}
\centerline{\psfig{figure=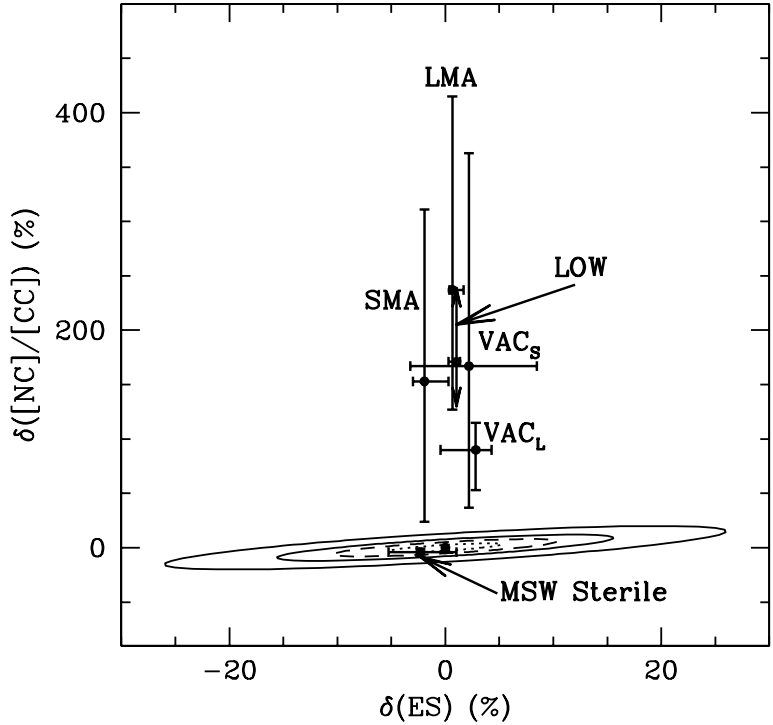,width=4in}} \tightenlines
\caption[]{\small The [NC]/[CC] double ratio versus the
neutrino-electron scattering double ratio.  The figure shows the
fractional difference in percent of the neutral current versus charged
current double ratio, [NC]/[CC], from the
no oscillation value of $1.0$, versus the fractional shift in percent
of the reduced neutrino-electron scattering ratio
from the Super-Kamiokande  value of $0.475$ [see Eq.~(\ref{eq:defnescminussk})].
Contours at $1\sigma,2\sigma,3\sigma$, and $5\sigma$ are shown
relative to the no-oscillation solution at $(0,0)$.
\label{fig:ncccvsesc}}
\end{figure}

Figure~\ref{fig:ncccvsesc} shows the predictions of the different
oscillation solutions in the $\delta{\rm [NC]/[CC]}$ versus $\delta
{\rm (ES)}$ plane.  Just as for Fig.~\ref{fig:ncccvsa} and
Fig.~\ref{fig:ncccvst}, all of the currently favored oscillation
solutions, with the exception of the MSW Sterile solution, are well
separated (more than $5\sigma$ away) from the standard model solution,
which lies at ($0.0,0.0$).  For some of the ${\rm VAC_S}$ solutions, the
predicted large positive value of $\delta {\rm [ES]}$ is incompatible
with, and hence distinguishable from, the predictions of the other
currently allowed solutions. This discrimination is a result of
combining the values of both [NC]/[CC] and [ES] since the measurement
of either of these parameters by itself would not permit, according to
Fig.~\ref{fig:ncccvsesc}, the isolation of these ${\rm VAC_S}$ solutions.

\subsection{Electron-scattering and CC double ratio versus CC energy
spectrum} 
\label{subsec:escccvst}

\begin{figure}{!t}
\centerline{\psfig{figure=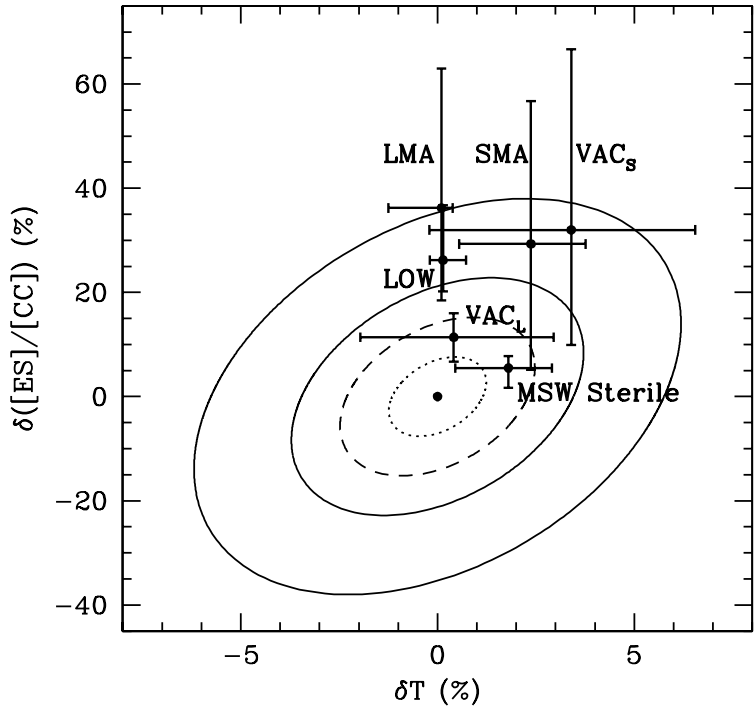,width=4in}}
\tightenlines
\caption[]{\small The [ES]/[CC] double ratio versus the average CC
electron recoil energy.   The
figure shows the fractional  difference in percent 
of the electron-scattering-CC double ratio, [ES]/[CC], 
from the no-oscillation solution of $ [ES]/[CC] = 1.0$, versus the fractional 
shift in percent of the average electron recoil energy, $\langle T\rangle$ 
(Sec.~\ref{sec:shapespectrum}), from the no-oscillation value of
Eq.~(\ref{eq:t0}).  Contours at $1\sigma,2\sigma,3\sigma$, and
$5\sigma$ are shown relative to the no-oscillation solution at
$(0,0)$.
\label{fig:escccvst}}
\end{figure}

Figure~\ref{fig:escccvst} shows the predictions of the different
oscillation solutions in the $\delta{\rm [ES]/[CC]}$ versus $\delta
T$ plane.  Although there are some predictions that extend well beyond
the $5\sigma$ contour in Fig.~\ref{fig:escccvst}, these outlying
predictions occur mostly for large values of [ES]/[CC] and should show
up directly by comparing the neutrino-electron scattering rate with
the CC (neutrino absorption) rate (see the discussion in
Ref.~\cite{snoshow}). The additional measurement of the first moment
of the CC distribution, $\langle T\rangle$, does not add much to the discriminatory
power of [ES]/[CC].  

\subsection{CC rate versus CC energy spectrum}
\label{subsec:ccratevst}

\begin{figure}{!t}
\centerline{\psfig{figure=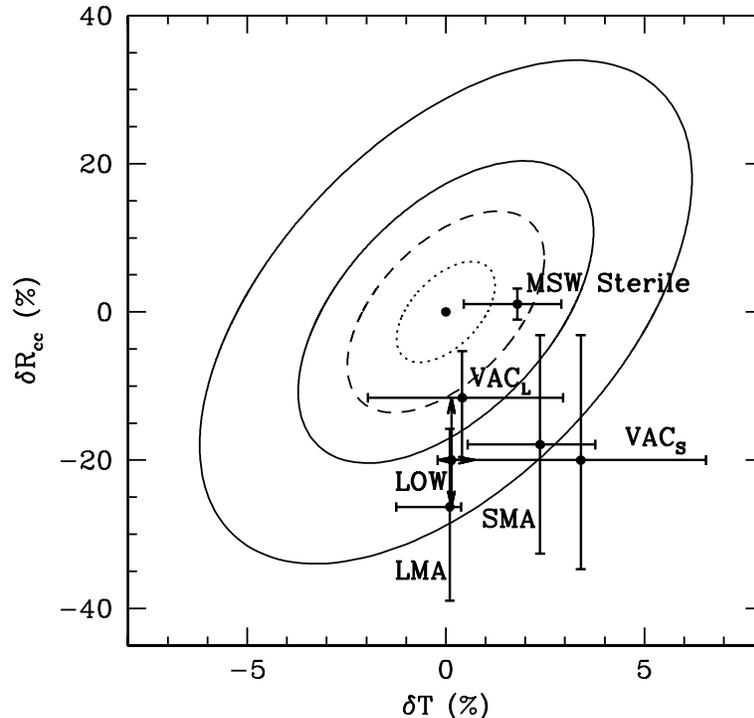,width=4in}}
\tightenlines
\caption[]{\small The CC rate versus the average recoil energy, $\langle T\rangle$.  The
figure shows the percent difference of the CC double rate, $R_{\rm
cc}$, discussed extensively in Ref.~\cite{snoshow}, from 
the no-oscillation solution of $R_{\rm cc} = 0.475$, versus the fractional
difference [see Eq.~(\ref{eq:defndeltat})] of the average 
electron recoil energy, $\langle T\rangle$, from the no-oscillation value of
Eq.~(\ref{eq:t0}).  Contours at $1\sigma,2\sigma,3\sigma$, and
$5\sigma$ are shown relative to the no-oscillation solution at
$(0,0)$.
\label{fig:rccvst}}
\end{figure}

Figure~\ref{fig:rccvst} displays the six currently favored oscillation
solutions in the plane of the CC rate, $R_{\rm CC}$, and the first
moment of the CC electron-recoil energy spectrum, $\langle T\rangle$. Some of the
currently allowed ${\rm VAC_S}$, SMA, and LMA solutions lie in this 
plane more than $5\sigma$ from the no-oscillation position. However,
there are also currently allowed oscillation solutions that fall
considerably closer to the no-oscillation point at ($0.0,0.0$).

\section{Summary and Discussion}
\label{sec:discussion}
We concentrate in this section on describing the results for the
predictions of the six currently favored neutrino oscillation
solutions that are globally consistent at the $99$\% C.L. with all of
the available neutrino data. The neutrino solutions are described in
Table~\ref{tab:bestfit} and Fig.~\ref{fig:survival}.

We begin this section by summarizing the results for parameters for
which the estimated uncertainties are relatively small: 1) the
neutral-current over charged current double ratio,
Sec.~\ref{subsec:discussnccc}; 2) the shape of the CC electron
recoil energy spectrum, Sec.~\ref{subsec:discussspectrum}; 3) the
day-night difference for the CC and for the NC,
Sec.~\ref{subsec:discussdaynight}; and 4) seasonal effects,
Sec.~\ref{subsec:seasonal}. Altogether, we discuss six measurable
quantities in
Sec.~\ref{subsec:discussnccc}--Sec.~\ref{subsec:seasonal}.

We summarize the principal uncertainties, theoretical and
experimental, in Sec.~\ref{subsec:discussuncertainties}.
The uncertainties due to the $hep$ flux and the neutrino interaction
cross sections are emphasized in this section; the estimates of the
experimental uncertainties are very preliminary.

We then describe the predicted values  and the potential
inferences from SNO measurements for the CC rate,
Sec.~\ref{subsec:discusscc}, for the NC rate,
Sec.~\ref{subsec:discussnc}, and for the neutrino-electron
scattering rate, Sec.~\ref{subsec:discussesc}.  Next we discuss in
Sec.~\ref{subsec:discussesctocc} the neutrino-electron scattering
to CC double ratio, which has some of the same advantages as the
neutral-current to charged current double ratio, but suffers from a
relatively large uncertainty in the CC interaction cross section.
Finally we summarize in Sec.~\ref{subsec:discusssmoking2} our
initial exploration of combining the analysis of different smoking gun
indicators of neutrino oscillations.

\subsection{Neutral current over charged current double ratio:
[NC]/[CC]}
\label{subsec:discussnccc}
All five of the currently favored oscillation solutions with active
neutrinos predict that the double ratio, [NC]/[CC], will be separated
from the no-oscillation value of $1.0$ by more than $9\sigma$,
estimated non-statistical errors.  The uncertainties due to the cross
sections and to the solar model almost cancel out of the double ratio.
The minimum predicted value for [NC]/[CC] is $1.24$ and the maximum
predicted value is $5.15$, all for a $5$ MeV CC threshold. The
estimated $1\sigma$ total non-statistical error is only $\pm 0.026$;
the statistical error will be the largest uncertainty unless more than
$5.5 \times 10^3$ NC events are detected.  The sterile neutrino
solution lies in a disjoint region of [NC]/[CC] from $0.92$ to $0.99$.

The results are summarized in Fig.~\ref{fig:double} and are given in more
detail, for two different thresholds of the CC electron recoil energy,
in Table~\ref{tab:doubleratio} of Sec.~\ref{sec:ncovercc}.

The double ratio [NC]/[CC] is an ideal smoking gun indicator of
oscillations into active neutrinos.

\subsection{The shape of the CC electron recoil energy spectrum:  \boldmath$\langle T\rangle$
and \boldmath$\sigma(T)$}
\label{subsec:discussspectrum}

The shape of the CC electron recoil energy spectrum can be
characterized by the first moment, $\langle T\rangle$, and the
standard deviation, $\sigma(T)$, of the electron kinetic energy,
$T$. With precision measurements of the spectrum, special features of
the recoil energy spectra may also be detectable
(cf. Fig.~\ref{fig:sixspectra}).

If there are no oscillations, the first moment is $\langle T\rangle_0 =
7.422\times (1 \pm 0.013) ~{\rm MeV}$, where the estimated uncertainty
includes both the measurement and the calculational uncertainties.
The best-estimate predictions for the different oscillation solutions
correspond to a fractional shift between $0.1$\% (LMA and LOW
solutions) to $3.7$\% (${\rm VAC_S}$ solution). The largest predicted
value of the shift is $7.5$\% (${\rm VAC_S}$ solution). The shift in
the first moment may be measurable for the SMA, ${\rm VAC_S}$, ${\rm
VAC_L}$, and MSW sterile neutrinos, but will be too small for a
definitive measurement if the LMA or LOW solutions are correct.  On
the other hand, only the ${\rm VAC_S}$ solutions predict that the
measured deviation of $\langle T\rangle$ from $\langle T\rangle_0$ will exceed three standard
deviations for as much as half of the currently allowed solution
space.

Figure~\ref{fig:deltat} and Table~\ref{tab:firstmoment} (of
Sec.~\ref{sec:shapespectrum}) show the predicted range of shifts in
the first moment, $\langle T\rangle$, of the recoil energy spectrum.  A measurement
of $\langle T\rangle$ to a $1\sigma$ accuracy of $\sim 100$ keV will significantly
reduce the allowed solution space for neutrino oscillations, but may
not uniquely favor one particular solution.

The calculated no-oscillation value of the standard deviation of the
CC recoil energy spectrum is 
$\sigma_0 = \langle\sigma^2\rangle^{1/2}_0
= 1.852(1 \pm 0.049)$ MeV.
It will be difficult to measure the predicted shifts
from $\sigma_0$ since the total spread in shifts given in
Table~\ref{tab:secondmoment} is  from $-29$ keV to $199$ keV,
while the estimated calculational and non-statistical measurement
uncertainties are $\pm 86$ keV.

\subsection{Day-night difference: \boldmath$A_{\bf N-D}$ and $A_{\rm N-D}({\rm NC})$ }
\label{subsec:discussdaynight}

In the absence of neutrino oscillations, there is a purely geometrical
day-night difference that we have defined as the No Oscillation (NO)
effect and whose value for the SNO detector we have given in
Eq.~(\ref{eq:snonooscdaynight}). We have removed the NO effect from
all of the calculated day-night values given in this paper. 

For the currently favored MSW active neutrino solutions, the best-fit
predictions for the average difference between the nighttime and the
daytime CC rates, $A_{\rm N-D}$, vary from $2$\% for the SMA solution
to $12.5$\% for the LMA solution, all for a $5$ MeV recoil electron
energy threshold.  Small values ($< 1$\%) of the day-night difference
would be consistent with any of the three MSW solutions, but very
large values of the day-night difference are only expected for some of
the LMA solutions. The maximum expected difference for the most
extreme LMA solution is $28.5$\%, whereas the maximum difference
expected for the LOW (SMA) solution is $13.5$\% ($12$\%).  The MSW
sterile solution predicts values for the day-night asymmetry between
$-0.5$\% and $+1.1$\%.  Table~\ref{tab:daynight} presents similar
results also for an $8$ MeV electron recoil energy threshold.

For vacuum oscillations, the predicted values of the day-night effect
are small, but non-zero (see Table~\ref{tab:daynight}).

Initially, the dominant uncertainty for the day-night effect will be
purely statistical.  The most difficult problems will ultimately arise 
from systematic effects, such as the symmetry of the detector and
the separation of the CC events from NC and scattering events, that
will have to be modeled by a detailed SNO Monte Carlo simulation.  
The purely statistical error may be of order $4$\% after one full year of
operation. Whether or not the systematic errors affect in an important
way the measurement of the day-night effect will  depend upon the
actual magnitude of $A_{\rm N-D}$ and the size of the systematic
uncertainties. 

Figure~\ref{fig:daynight} and Table~\ref{tab:daynight} present the
numerical results for the CC day-night asymmetry which is defined by
Eq.~(\ref{eq:daynightdefn}) of Sec.~\ref{sec:daynight}. 
If one of the MSW active neutrino solutions is correct, then the
day-night difference could become apparent early in the operation of
SNO. This possibility exists for the MSW active solutions, but is not
required. 

There is no day-night effect in the NC for oscillations into active
neutrinos. Oscillations into sterile neutrinos give a small effect:
$A_{\rm N-D}({\rm NC,~MSW~Sterile}) ~=~-0.001^{+0.006}_{-0.002}$ [see
Eq.~(\ref{eq:dnsterile})].  This effect is important in principle, since
it is a clear distinction between active neutrinos and sterile
neutrinos.  However, the predicted size is too small to be measured
with SNO.

All of the currently favored neutrino oscillation solutions (MSW or
vacuum oscillations into active neutrinos, as well as MSW oscillations into
sterile neutrinos), predict that 

\begin{equation}
\mid A_{\rm N-D}({\rm NC})\mid ~<~ 0.01.
\label{eq:discussa}
\end{equation}
We have written Eq.~(\ref{eq:discussa}) in its most general form. Of
course, $A_{\rm N-D}({\rm NC})$ is predicted to be identically zero for
all neutrino oscillations (vacuum or MSW) into active neutrinos.

The measurement by SNO of $A_{\rm N-D}({\rm NC})$ will be an important
test of neutrino oscillation models. If we obtain independent evidence
that solar neutrino oscillations involve active neutrinos, or if one
hypothesizes that sterile neutrinos play no role in solar neutrino
oscillations, then the measurement of $A_{\rm N-D}({\rm NC})$ can be
regarded as a test of the standard electroweak model.

\subsection{Seasonal effects: \boldmath$A_{\bf W-S}$}
\label{subsec:seasonal}

Figure~\ref{fig:wintersummer} and Table~\ref{tab:wintersummer} give
the amplitudes of the CC winter-summer differences, $A_{\rm W-S}$, that
are predicted by the favored neutrino oscillation solutions. The
results can be compared with the amplitudes expected from the orbital
motion of the earth, which are given in Eq.~(\ref{eq:orbitalshort})
and Eq.~(\ref{eq:orbitallong}). In all cases, the current best-fit
oscillation solutions predict a winter-summer amplitude that is less
than the amplitude due to the earth's orbital motion. Only for a small
fraction of the currently allowed oscillation parameters does the
predicted amplitude due to oscillations exceed the amplitude due to
the earth's orbital motion.

We conclude that it will probably be difficult to measure $A_{\rm
W-S}$.  However, we note that the prediction that $A_{\rm W-S}$ is
small is a prediction that can and should be tested.

\subsection{Uncertainties}
\label{subsec:discussuncertainties}
Table~\ref{tab:uncertainties} presents a convenient summary of the
estimated calculational and measurement uncertainties for different
experimental quantities that will be determined by SNO.  We present in
Sec.~\ref{sec:uncertainties} a full description of how we estimate
these uncertainties. We do not include the effects of background
events; there is no reliable way of estimating the background prior to
actual measurements in the SNO detector. We also do not include
misclassification uncertainties, e.g., ES events mistaken for CC
events or NC events mistaken for CC events. These errors must be
determined by the detailed SNO Monte Carlo simulations.

The quantitative influence of the $hep$ flux of
neutrinos on the measurement accuracy of different quantities has been
evaluated here for the first time. In addition, we include estimates
of uncertainties due to the width of the resolution function for the
recoil electron energies, the absolute energy scale, the $^8$B
neutrino energy spectrum, the interaction cross sections, and the number
of events counted.

Our present limited experimental knowledge of the $hep$ flux causes an
uncertainty of $\sim 2$\% in all three of the rates that will be
measured by SNO, i.e., the CC rate, the NC rate, and the
neutrino-electron scattering rate (see
Table~\ref{tab:uncertainties}). However, measurements of the CC
spectrum by SNO can reduce the uncertainty in the $hep$
flux~\cite{snoshow} and therefore decrease the contribution of the
$hep$ to the error budgets of different SNO measurables. Most
recently, an improved theoretical calculation~\cite{marcucci} of the
low energy cross section factor for the $hep$ reaction has increased
the best-estimate of the flux from the standard solar model by about a
factor of five relative to the value given in BP98 and used in the
present paper.

The neutrino cross section uncertainties for the CC and the NC rates
are the largest entries in Table~\ref{tab:uncertainties}.  The lack of
knowledge of these interaction cross sections will limit the
interpretation of the measured rates of both the CC and the NC
interactions. 

Table~\ref{tab:crosssections} summarizes the recent cross section
calculations for the CC, NC, and for the ratio, NC/CC. We have
computed uncertainties in measurable quantities using the entries in
Table~\ref{tab:crosssections} and the algorithm given in
Eq.~(\ref{eq:crossuncertain}).  It is a matter of judgment as to how
many standard deviations should be assigned to the difference computed
from Eq.~(\ref{eq:crossuncertain}).  We believe that we are being
reasonable and conservative in regarding this difference as $1\sigma$.
But, we stress the need to greatly increase the limited number of
entries in Table~\ref{tab:crosssections} so that a more informed
estimate of the uncertainties can be made. The need for additional
calculations is particularly urgent for the double reduced ratio
[NC]/[CC].  We have used just the two entries in
Table~\ref{tab:crosssections} to estimate that the uncertainty in
[NC]/[CC] is an order of magnitude less than the separate
uncertainties in [NC] and [CC] . While plausible, it is essential to
check that this cancellation of uncertainties in the double ratio is
indeed a general characteristic of accurate calculations of the
neutrino interaction cross sections.

Towner~\cite{towner} has computed radiative corrections for both the
CC and the NC neutrino reactions on deuterium. The effect of the
radiative corrections is generally small. Radiative corrections change
the first moment by about $0.1$\% and the second moment by about
$0.3$\%. Although not computed by Towner, the effect of radiative
corrections on time-dependent quantities such as the day-night effect,
the zenith angle distribution, and the seasonal effects is expected to
be similarly small ($\leq 0.3$\%). For the double ratio [NC]/[CC], the
effect of the radiative corrections is larger, $0.5$\%, if the photons
from the inner bremsstrahlung in the CC reactions are not detected. In
the extreme case in which all of the inner bremsstrahlung photons are
somehow detected by SNO, the [NC]/[CC] ratio would be increased by
$4$\%
\footnote {For neutrino-electron scattering, radiative corrections are
also small and have been computed explicitly by Bahcall, Kamionkowski,
and Sirlin~\cite{sirlin}.  The Super-Kamiokande collaboration includes
these calculated radiative corrections in their analyses~\cite{superk}.}.

The knowledge of the cross section uncertainties can be improved by
further calculations, especially those based upon chiral
symmetry. Calculations should be carried out for a variety of models
and approximations and with the full range of allowed nuclear and
particle physics parameters. An initial step in this direction has
been taken by Butler and Chen~\cite{bc}, who have calculated the NC
reaction in effective field theory. A full exploration of the allowed
range of CC and NC cross sections for neutrinos incident on deuterium
is an urgent and important task for the theoretical nuclear physics
community.  

Further experimental work on neutrino interactions with deuterium
would be extremely valuable.  Butler and Chen~\cite{bc} have pointed
out that a measurement of the two-body matrix element could be made
using the reaction $e$ + $^2$H $\rightarrow e + n + p$. This is an
urgent and important task for the experimental nuclear physics
community.  More precise measurements of the anti-neutrino
disintegration cross sections made with reactors would be valuable in
choosing between and guiding theoretical calculations (for a
state-of-the-art discussion of the experimental possibilities see
Ref.~\cite{riley}).  It would also be useful to test the accuracy of
the calculational procedures, albeit at higher neutrino energies, by
performing neutrino absorption and disassociation experiments on
deuterium with a stopped muon beam.

\subsection{The CC rate: [CC]}
\label{subsec:discusscc}
The charged current rate will be one of the first results to be
obtained with SNO. The reduced neutrino-absorption rate, [CC], defined
by Eq.~(\ref{eq:defnccratio}), can be compared with the
neutrino-electron scattering ratio measured by
Super-Kamiokande~\cite{superk}, $0.475 \pm 0.015$. If the CC ratio is
measured to be less than $0.475$, then that would be evidence that
$\nu$-$e$ scattering includes contributions from muon or tau neutrinos
and therefore neutrino oscillations are occurring.

Table~\ref{tab:ccratio} presents the predicted CC reduced rates,
[CC], for the six currently favored oscillation solutions. For
example, the predicted ratio for oscillations into active neutrinos
ranges from $0.29$ (LMA, minimum value) to $0.46$ (SMA, maximum
value), if the recoil electron energy threshold is set at $5$ MeV.
There is about an equal chance, according to Table~\ref{tab:ccratio}
and Fig.~2 of Ref.~\cite{snoshow}, that the measured value of [CC]
will lie more than $3\sigma$ below the no-oscillation value of $0.475$. 
Four of the solutions, the LMA, SMA, LOW, and ${\rm VAC_S}$ solutions, all
have best-fit global solutions that predict [CC] $< 0.40$ and each of
these solutions has some region of neutrino parameter space that gives
values as low as $0.35$ or below. 

The MSW sterile solution predicts values for [CC] that are very close
to $0.475$ for an electron recoil energy threshold of
$5$ MeV. For a threshold of $8$ MeV, the MSW sterile solution predicts
values for [CC] that even exceed $0.475$ (see
explanation of this interesting fact in Ref.~\cite{snoshow}).

The discriminatory power of the CC rate measurement could be increased
significantly if the uncertainty in the CC cross section could be
decreased (see discussion in Sec.~\ref{subsec:discussuncertainties}
above).

\subsection{The NC rate: [NC]}
\label{subsec:discussnc}

The reduced neutral current rate, [NC], should be equal to 1.0 if the
standard solar model prediction of the $^8$B flux is exactly correct
and if there are no oscillations into sterile neutrinos. Oscillations
into active neutrinos would preserve the neutral current rate and
would not change the 1.0 predicted value of the reduced rate.

The interaction cross section constitutes the largest uncertainty,
$\pm 6$\%, in determining experimentally the reduced neutral current
rate. The uncertainties in calculating the solar flux  ($+ 18$\%,
$-16$\%, see Ref.~\cite{bp98}) provide the biggest complication in
interpreting the neutral current measurement directly in terms of
neutrino physics [see Eq.~(\ref{eq:ncstandard}) for the uncertainties
in measuring and interpreting [NC]].

The sensitivity of [NC] to the true solar flux is a problem for
particle physics, but an advantage for astrophysics. 

The measurement of the neutral current rate will provide crucial
information about the true $^8$B solar neutrino flux provided there
are no oscillations into sterile neutrinos.  Unfortunately, the MSW
sterile neutrino prediction, ${\rm [NC]_{Sterile} = 0.48 \pm 0.01}$
 [see Eq.~(\ref{eq:ncsterile})], is within about $3\sigma$ of the
no-oscillation value of $1.0$ when one includes the uncertainty in the
solar model flux.  Hopefully, the uncertainty in the predicted value
of the standard solar model flux will be reduced somewhat by precise
laboratory measurements of the $^8$B production cross section that are
now in progress.

A measurement of [NC] larger than or close to $1.0$ would be evidence
against sterile neutrino oscillations and would support the solar
model estimate for the $^8$B flux (provided the experimental value
is not larger than $1.5$).  A measurement between $1.0$ and $0.5$
could be interpreted as indicating a solar flux somewhat lower than
the best estimate or as providing evidence for sterile neutrinos.  A
measurement significantly below $0.5$ would be a clear indication of
oscillations into sterile neutrinos, but would conflict with the
Kamiokande~\cite{kamiokande} and Super-Kamiokande~\cite{superk}
measurements of the $\nu$-$e$ scattering rate.

Bilenky and Giunti~\cite{bg} have pointed out that a comparison of the
time-dependence of the CC and NC SNO rates on the time scale of the
$11$year solar cycle could test the spin-flavor precession
scenario. According to this hypothesis, the NC rate would
remain constant throughout the solar cycle while the CC rate would
vary with phase in the cycle.

\subsection{The neutrino-electron scattering rate: [ES]}
\label{subsec:discussesc}

Table~\ref{tab:escratio} and Fig.~\ref{fig:esc} show the predicted
values of the reduced neutrino-electron scattering ratio, [ES], for
the six currently favored oscillation solutions. Not surprisingly, the
predicted ratios cluster close to the Super-Kamiokande value of
$0.48$~\cite{superk}. The most extreme values range from $0.45$
 (minimum allowed for the MSW sterile solution) and $0.52$ (maximum
allowed for the ${\rm VAC_S}$ solution).  These values are all for a recoil
electron energy threshold of $5$ MeV (see Table~\ref{tab:escratio} for
results for an $8$ MeV threshold) and a $99$\% C.L. for the allowed
range of oscillation solutions.

For the first five or ten years of operation of SNO, the dominant
measurement uncertainty for [ES] will be statistical
 (cf. Table~\ref{tab:escratio}). The observed rate of neutrino-electron
scattering events in SNO is expected to be only $\sim 10$\% of the CC
rate.

Although the SNO detector is different from either the Kamiokande or
the Super-Kamiokande detectors, and the value of [ES] depends somewhat
on threshold (see Table~\ref{tab:escratio}) and on the instrumental
parameters such as the width of the energy resolution function, the
bottom-line results for [ES] should be similar in all cases for these water
Cherenkov detectors. 

The SNO measurement of $\nu$-$e$ scattering will
provide an important check of SNO versus Kamiokande and
Super-Kamiokande and vice-versa.  In addition, the value of [ES] as
determined in SNO can be used in connection with other SNO
measurements to constrain the allowed neutrino parameter oscillation
space.

\subsection{The \boldmath$\nu$-$e$ to CC double ratio: [ES]/[CC]}
\label{subsec:discussesctocc}

The double ratio of [ES]/[CC] is, like the [CC]/[NCC] double ratio,
largely insensitive to solar model uncertainties (see
Table~\ref{tab:uncertainties}). Moreover, some of the systematic
measurement uncertainties are reduced because the same techniques are
used to detect $\nu$-$e$ scatterings ([ES]) and neutrino absorption
 ([CC]).  The principal difficulty in interpreting measurements of the
double ratio [ES]/[CC] at the present time is the large uncertainty,
$5.8$\%, in the CC reaction cross section.  This uncertainty is almost
six times larger than any other known contributor to the [ES]/[CC]
error budget  [see Eq.~(\ref{eq:escccstandard})].

Figure~\ref{fig:escdouble} and Table~\ref{tab:escdoubleratio}
present the predicted range of values for [ES]/[CC] for the six
currently favored neutrino oscillation solutions.  For oscillations
into active neutrinos, $1.03 < {\rm [ES]/[CC] } < 1.65$ for a $5$ MeV
recoil electron energy threshold. The corresponding limits are $1.05$
and $1.67$ for an $8$ MeV energy threshold (see
Table~\ref{tab:escdoubleratio}). The total non-statistical
uncertainty is estimated to be $7$\%  [see Eq.~(\ref{eq:escccstandard})]
and the statistical uncertainty will be about $5$\% after the
accumulation of  $5000$ CC events.

We conclude that Nature has adequate opportunity to choose an
oscillation solution into active neutrinos in which the ratio
[ES]/[CC] is many sigma from the no-oscillation value of $1.0$.
Nevertheless, the contrast between the no-oscillation value and the
currently favored oscillation predictions is much less for [ES]/[CC]
than it is for [NC]/[CC]. The greater power of [NC]/[CC] can be seen
most clearly by comparing Fig.~\ref{fig:double}, which has a vertical
scale that extends from $0.5$ to $7.0$, with Fig.~\ref{fig:escdouble},
which has a vertical scale that extends only to $3.0$.

\subsection{Smoking gun versus smoking gun}
\label{subsec:discusssmoking2}

The full diagnostic power of SNO will be achieved by analyzing
simultaneously all of the measurements,  including 
upper limits. This
full analysis requires detailed and mature Monte Carlo simulations
based upon experimental calibrations. 

Figure~\ref{fig:ncccvsa}--Figure~\ref{fig:rccvst} provide an
illustrative introduction of what can ultimately be achieved by the
simultaneous analysis of the full set of SNO measurables. The figures
are two dimensional slices in the multi-dimensional SNO parameter
space; we plot one smoking gun against another smoking gun. Contours
ranging from $1\sigma$ to $5\sigma$ are shown for the no-oscillation
case and include estimates for the error correlations. The error bars
for the different oscillation scenarios represent the range of values
predicted for each smoking gun independently.

\section{What are our most important conclusions?}
\label{sec:mostimportant}

The paper contains many specific results. Here is our personal list of
our most important conclusions.

(1) {\bf The neutral current to charged current double ratio.} All
    currently favored active neutrino oscillation solutions predict a
    value for the double ratio, [NC]/[CC], of neutral current to
    charged current event rates that is, with our best estimates for
    the theoretical and experimental uncertainties, more than
    $9\sigma$ away from the no-oscillation solution (neglecting
    statistical uncertainties).  If statistical uncertainties are
    included for $5000$ CC events (and $1219$ NC events), then the
    minimum discrepancy is reduced to $6\sigma$.

(2) {\bf Day-Night differences in the CC rate.} Large differences are
    predicted between the day and the night CC rates for some
    currently favored MSW solutions.  For a $5$ MeV electron recoil
    energy threshold, the best-fit differences vary between $-0.1$\%
    (MSW Sterile) and $12.5$\% (LMA). The largest predicted value
    among all the currently allowed solutions is $28.5$\% (LMA), which
    could be detectable in the first year of operation of SNO. Similar
    results are predicted for an $8$ MeV recoil energy
    threshold. Vacuum oscillations have average day-night differences
    of order $1$\% or even less. Small values ($< 1$\%) of the CC
    day-night rate difference would be consistent with any of the six
    currently favored two-flavor oscillation solutions.  The day-night
    difference of the NC is predicted to be $< 0.01$\% for all
    solutions (and is non-zero for the MSW Sterile solution).

(3) {\bf Uncertainties.} The uncertainties in the absolute values of
    the neutrino cross sections for the CC and for the NC current
    reactions are the largest known uncertainties.  These
    uncertainties limit the interpretation of the separate CC and NC
    rates, but cancel out (to an accuracy of better than $1$\%) of the
    [NC]/[CC] ratio.

(4) {\bf  Specrum distortion.} 
The first moment of the CC electron recoil energy spectrum describes
well the predicted deviations for all except the ${\rm VAC_L}$ solution.
For all the MSW solutions and for the ${\rm VAC_S}$ solution, the predicted
spectrum distortion is smooth and monotonic in the region accessible
to SNO (see Fig.~\ref{fig:sixspectra}) and hence can be characterized
by a single parameter. The currently favored oscillation solutions
predict a range of deviations of the first moment, most of which are
less than the estimated $3\sigma$ experimental uncertainty. The
largest predicted deviations are for the ${\rm VAC_S}$ (best-fit predicted
deviation $283$ keV) and SMA (best-fit predicted deviation $218$ keV) 
solutions.

The ${\rm VAC_L}$ solution generically predicts a bump and a dip in the low
and middle energy region of the SNO electron recoil energy spectrum (see
Fig.~\ref{fig:sixspectra}). For the best-fit ${\rm VAC_L}$ solution, this
modulation is about $30$\% and if observed would be strong evidence
for the ${\rm VAC_L}$ scenario. The predicted modulation occurs in a region
where the event rate is expected to be relatively high.

(5) {\bf Characteristic size of effects.} The current best-fit global
    neutrino oscillation solutions typically predict small effects, of
    order several percent or less, for all of the quantities that are
    sensitive to oscillations which SNO will measure, except
    [NC]/[CC].  However, for some allowed oscillation solutions, the
    difference between the day and the night rates and the distortion
    of the shape of the CC electron recoil energy spectrum may be
    relatively large.

(6) {\bf Sterile neutrinos.} The current best-fit MSW Sterile solution
    predicts, relative to the no-oscillation solution, a $1.8$\% shift
    in the first moment of the CC electron recoil energy spectrum and
    a CC rate that is larger than for the other currently allowed
    oscillation solutions. The neutral current rate is predicted to be
    $0.465 \pm 0.01$ of the standard solar model rate, i.e., the MSW
    Sterile solution predicts a much smaller value for the neutral
    current rate than the other allowed oscillation solutions. The
    sterile solution also predicts a small but non-zero value for the
    difference between the NC rate during the day and the NC rate at
    night. The CC day-night difference is predicted to be small, but
    not as small as for the NC day-night difference. It will be
    difficult to discriminate with SNO between the no-oscillation
    solution and the currently allowed MSW Sterile solution.

The Sudbury Neutrino Observatory will enrich particle physics with
measurements of many effects that are sensitive, in
different ways, to neutrino oscillations .    

We are grateful to colleagues in the SNO collaboration who have by
their important experimental work and by their stimulating comments
raised the questions that this paper addresses.  We are indebted to
E. Akhmedov, E. Beier, S. Bilenky, D. Cowan, E. Kearns, J. Feng,
M. Fukugita, K. Kubodera, E. Lisi, A. McDonald, and Y. Nir for
valuable comments on a draft copy of this manuscript. JNB and AYS
acknowledge partial support from NSF grant No. PHY95-13835 to the
Institute for Advanced Study and PIK acknowledges support from NSF
grant No. PHY95-13835 and NSF grant No. PHY-9605140.

\end{document}